\documentclass[12pt,epsf]{article}
\usepackage{graphics}

\makeatletter

\newcommand{\LyX}{L\kern-.1667em\lower.25em\hbox{Y}\kern-.125emX\@}
\newcommand{\noun}[1]{\textsc{#1}}

\newcommand{\lyxaddress}[1]{
  \par {\raggedright #1 
  \vspace{1.4em}
  \noindent\par}
}

\usepackage{graphics}


\makeatother

\begin{document}

\title{Issues on Orientifolds: On the brane construction of gauge theories with SO(2n)
global symmetry}

\author{Amihay Hanany\protect\( ^{1}\protect \) and Alberto Zaffaroni\protect\( ^{2}\protect \)}

\maketitle

\lyxaddress{\protect\( ^{1}\protect \)Center for Theoretical Physics, Laboratory for Nuclear
Science, Department of Physics, Massachusetts Institute of Technology, Cambridge,
Massachusetts 02139, U.S.A.\\
 hanany@mit.edu}

\lyxaddress{\protect\( ^{2}\protect \)Theory Division, CERN, Ch 1211 Geneva 23, Switzerland\\
alberto.zaffaroni@cern.ch}

\begin{abstract}
We discuss issues related to orientifolds and the brane realization for gauge
theories with orthogonal and symplectic groups. We specifically discuss the
case of theories with (hidden) global \( SO(2n) \) symmetry, from three to
six dimensions. We analyze mirror symmetry for three dimensional N=4 gauge theories,
Brane Box Models and six-dimensional gauge theories. We also discuss the issue
of T-duality for \( D_{n} \) space-time singularities. Stuck D branes on \( ON^{0} \)
planes play an interesting role.
\end{abstract}
\vfill

\begin{flushright}

CERN-TH/99-82\\
 MIT-CTP-2845\\
 hep-th/9903242

\end{flushright}

\eject \tableofcontents

\section{Introduction}

This paper discusses issues related to orientifolds. The motivation which led
to this project was to understand various strong coupling behaviour of orientifold
planes. These planes play an important role in understanding the non-perturbative
properties of string theories. They can also be used for realizing and studying,
using branes, gauge theories with orthogonal and symplectic gauge groups; in
this perspective, the dictionary for translating different gauge theories into
brane configurations is still under construction. There are several types of
orientifold planes. For most of them, we know, at least classically, what is
the gauge theory realized on branes living in the presence of these orientifolds.
However, as will be discussed below, there are cases (the somehow exotic plane
\( \hat{O} \) \footnote{
See the next few paragraphs for a discussion on this object. 
} or the \( ON \) planes acting on NS-branes) in which the entries in the dictionary
are still mysterious. Moreover, even when everything is understood at the classical
level, the strong coupling behaviour of these configurations is not well understood;
this prevents us from extracting non-perturbative results from the brane configuration
for the corresponding gauge theory. Unfortunately, this lack of understanding
is common to many different orientifold configurations.

In this paper, we make a first step in the direction of better understanding
these configurations and we analyze a class of theories, with various supersymmetries
and in various dimensions, with global \( SO(2n) \) symmetry. The issues which
will be discussed here in more detail are:

\begin{itemize}
\item Mirror Symmetry in three dimensions.
\item Various constructions of Brane Box Models involving orientifold planes. 
\item Some issues in six dimensional theories. 
\item More constructions of Finite four dimensional gauge theories using branes. 
\end{itemize}
All the theories presented have in common a \( D_{n} \) type global symmetry.
Since, besides the use of an orientifold, theories with \( D_{n} \) global
symmetry can be realized by engineering space-time singularities, there is a
natural question that arises in relation to the models discussed in this paper
and that we will discuss in some detail: 

\begin{itemize}
\item What happens to a \( D_{n} \) space-time singularity when we perform a T-duality? 
\end{itemize}
The answer for this question in the case of a \( A_{k} \) singularity is extensively
discussed in the literature, and here we present the \( D_{n} \) case. The
understanding of the behaviour of singularities under T-duality helps in relating
(and therefore in better understanding) different approaches with branes to
the same gauge theory. For each of the theories considered in this paper, we
will discuss the fate of the configurations under T-duality and the relation
with different approaches in the literature.

Let us briefly summarize what is known about the orientifold zoo. 

There are several types of orientifold planes. The most familiar and most discussed
one is the \( Op \) plane which carries negative RR charge, that is in units
in which the charge of a Dp-brane is positive (we will take the charge of a
physical brane to be +1 in this paper). The charge of this \( Op \) plane is
\( -2^{p-5} \). A collection of \( n \) physical Dp-branes located at the
\( Op \) plane give rise to a \( D_{n} \) gauge theory on the world volume
of the Dp-branes, that is an enhanced \( SO(2n) \) gauge theory. Henceforth
this orientifold will be called \( Op^{-} \), in short. If there are \( 2^{p-5} \)
physical Dp branes located on top of an \( Op^{-} \) plane, we get a special
case in which the RR charge of this object is zero. The corresponding RR field
has no sources coming from such an object which makes it special and leads to
interesting phenomena in various dimensions and supersymmetries. For a recent
discussion for the case of \( O6 \) planes see \cite{asadjosh}. We will call
this object \( Op^{0} \). 

Another type of orientifold is also fairly discussed in the literature. This
orientifold plane carries an opposite charge with respect to the \( Op^{-} \)
plane, \( +2^{p-5} \), and will be called \( Op^{+} \). A collection of \( n \)
physical Dp-branes sitting on top of this orientifold plane give rise to an
enhanced \( C_{n} \) gauge theory on the world volume of the Dp-branes, that
is a \( Sp(n) \) gauge theory. 

With this realization of gauge theories, the natural question which arises is
what is the realization of \( B_{n} \) type gauge theories. (We remind that
\( A_{n} \) gauge theories are given by collecting \( n+1 \) branes on top
of each other, with no presence of an orientifold plane). The answer to this
question is also well known. If one puts a stuck Dp-brane on the \( Op^{-} \)
plane and, in addition, \( n \) physical branes, one gets the desired construction.
The charge of a stuck Dp-brane is \( +1/2 \), this means that the charge of
the orientifold with a stuck Dp-brane on it is \( 1/2-2^{p-5} \). Henceforth,
this orientifold plane will be called \( \widetilde{Op} \). 

A last type of orientifold plane is the less familiar object out of all other.
This orientifold plane was discussed in various papers \cite{witten,uranga,hori}
where the special cases of \( p=3 \) and \( p=4 \) were discussed. T-duality
suggests the existence of this orientifold plane for any \( p \) and this issue
will be discussed elsewhere\footnote{
It is interesting to look at configurations with orientifold planes and stuck
5 branes. The nature of the orientifold plane changes as one crosses the 5 branes
and strong coupling duals of such configurations lead to interesting results. 
}. This orientifold plane will be called \( \widehat{Op} \). A universal charge
formula for this orientifold plane is not clear at the moment.

Once we accept the existence of these types of orientifold planes, the next
natural question which comes to mind is what is the strong coupling behavior
of such planes? In many cases the answer appears in the literature. One notable
work on this issue is of Sen, \cite{Sen} where he gives a description of strong
coupling behaviour of \( Op^{-} \) planes for various values of \( p \). Less
is known about the other types of orientifolds. 

A special case is \( p=5 \). We ask what is the strong coupling of an \( O5 \)
plane. We will call such an object an \( ON \) plane which comes from the fact
that S-duality of Type IIB implies that this orientifold carries magnetic charge
with respect to the NS two-form of Type IIB. T-duality also implies that there
will be a similar object in Type IIA theory, that is an \( ON \) plane which
carries a NS two-form charge. As for the \( Op \) planes, we assume that there
are four types of such objects which will be denoted by \( ON^{-},ON^{+},\widetilde{ON},\widehat{ON} \).
The charges of these objects are \( -1 \) for \( ON^{-} \), \( +1 \) for
\( ON^{+} \), \( +\frac{1}{2} \) for \( \widetilde{ON} \) and not clear for
\( \widehat{ON} \). Similar to the \( Op^{0} \) object, we can have a configuration
with a physical NS brane sitting on top of an \( ON^{-} \) plane. This configuration
carries no NS two-form charge and will be denoted \( ON^{0} \). Like the \( Op^{0} \)
objects the absence of NS two-form charge leads to interesting field theory
consequences which makes this configuration special. 

Another set of questions which arise in the presence of such orientifolds is
what is the gauge theory which lives on the world volume of Dp-brane which have
some directions which are not parallel to the orientifold planes. As an example,
we would like to know what is the world volume gauge theory of a collection
of D4-branes which are sitting transverse to an \( \widetilde{ON} \) plane
in a supersymmetric fashion. All combinations of this type are of natural interest. 

In this paper we will mainly focus on theories with \( SO(2n) \) global symmetry.
The recent understanding of the properties of the \( ON^{0} \) plane \cite{senS,kapu}
helps us in extending previous analysis of mirror symmetry in three dimensions
\cite{hw,pz}, Brane Box Models \cite{hzfour, hstrassu, hu} and six-dimensional
theories \cite{karch, hzsix, karchbr} to the case with \( D_{n} \) symmetry.
Essentially, this paper extends what was done in the above mentioned papers
from the \( A_{k} \) series to the \( D_{n} \) series. In this context, it
is of great importance to understand what happens to a \( D_{n} \) ALE space
under T-duality. This issue has its own importance even besides its use for
realizing gauge theories with branes and it will be discussed in detail in this
paper. It is natural to ask about theories with \( B_{n} \) and \( C_{n} \)
global symmetry. This is related to the understanding of strong coupling behavior
of \( \widetilde{Op} \) and \( Op^{+} \) planes. This will be discusses elsewere.

The paper is organized as follows. In section \ref{onplane} we review and extend
the analysis of the \( ON^{0} \) plane properties. We extensively discuss the
behaviour of D-brane probes near the plane. This section contains the technical
tools that will be needed for all the examples in this paper. The subsection
\ref{section: oointer} is more technical; since the results in this subsection
will be used only for a particular pair of mirror theories and for six-dimensional
theories, the reader not interested in these topics may skip this subsection.
In section \ref{T} we discuss the T-duality for an ALE space of type \( D_{n} \).
In the following sections we discuss various examples of gauge theories in various
dimensions. In section \ref{mirror} we present several examples of mirror pairs
in N=4 three dimensional gauge theories. In section \ref{fourd} examples of
Brane Box Models and finite N=1 gauge theories in four dimensions. In section
\ref{section: sixd} examples of six-dimensional superconformal fixed points
and small instanton theories. We tried to keep each different section as self-contained
as possible, in such a way that the reader only interested in a particular class
of gauge theories or only interested in T-duality for ALE spaces may skip what
does not interest him. For each class of gauge theories we explicitly discuss
T-dual descriptions and try to make contact with different approaches. 

\def\nsp{NS\('\)}

\section{The \protect\( ON^{0}\protect \) Plane\label{onplane}}

A particularly important object for our purpose is the plane \( ON^{0} \),
obtained as a superposition of an \( ON^{-} \) plane and a physical NS brane.
It arises both as the strong coupling limit of \( O5^{-} \) planes and as an
essential ingredient in the T-duality for \( D \) singularities. 

Its importance is increased by the fact that it has a perturbative description
\cite{senS}. The \( ON^{0} \) plane can be considered as the fixed plane of
the perturbative \( (-1)^{F_{L}}R \) projection in a Type II string theory.
This orbifold projection is the combination of a space-time \( Z_{2} \) inversion
in four directions, \( R, \) and the operator \( (-1)^{F_{L}} \), the left-handed
space-time fermion number. There are fields living on the world-volume of \( ON^{0} \).
They are the twisted states of the orbifold projection, which, both in Type
IIA and Type IIB, have the same massless field content as a NS-brane\footnote{
The twisted states are vector supermultiplets of a \( (1,1) \) six-dimensional
theory in Type IIB and \( (2,0) \) tensor supermultiplets in Type IIA. 
}. 

\( ON^{0} \) acts as a sort of \textit{orientifold} projection for NS-branes.
\( n-1 \) physical NS-branes on top of the \( ON^{0} \) plane realize an \( SO(2n) \)
symmetry; this is the standard gauge symmetry of the system of NS-branes in
Type IIB, or, the non-Abelian symmetry acting on the tensionless strings of
the \( (2,0) \) theory in Type IIA. Notice that, in this description, one of
the Cartan generators of \( SO(2n) \) is not associated with the position of
one of the NS branes but rather with the twisted states of the orbifold projection\footnote{
It is interesting to consider the brane realization of the BPS spectrum of these
theories. The W-bosons are realized by D-strings, and the magnetically charged
two-branes are realized by D3 branes. Near the \( ON^{0} \), the realization
is different than the usual one near \( Op \) planes. See figure \ref{LAST}
for some of these states.
}. 

There are two properties of \( ON^{0} \) that make it special. They also explain
the proposed \( SO(2n) \) symmetry: 

\begin{itemize}
\item S duality. In the Type IIB context, \( n-1 \) NS-branes near \( ON^{0} \)
plane give the S-dual realization of \( (1,1) \) \( SO(2n) \) six-dimensional
theories with \( n \) physical D5-branes on top of an orientifold plane. S-duality
indeed maps D5-branes into NS-branes, and \( O5 \) planes into \( ON \) planes.
\( ON^{0} \) is the strong coupling limit of the corresponding \( O5^{0} \)
plane, the superposition of an \( O5^{-} \) orientifold and a physical D5 brane
\cite{senS}. As noticed in \cite{senS}, the \( O5^{0} \) plane, as opposed
to \( O5^{-} \) supports fields on its world-volume and it is not charged under
the RR six-form; it has therefore the right characteristic for being the strong
coupling limit of the orbifold fixed plane.
\begin{figure}
{\par\centering \resizebox*{8cm}{!}{\includegraphics{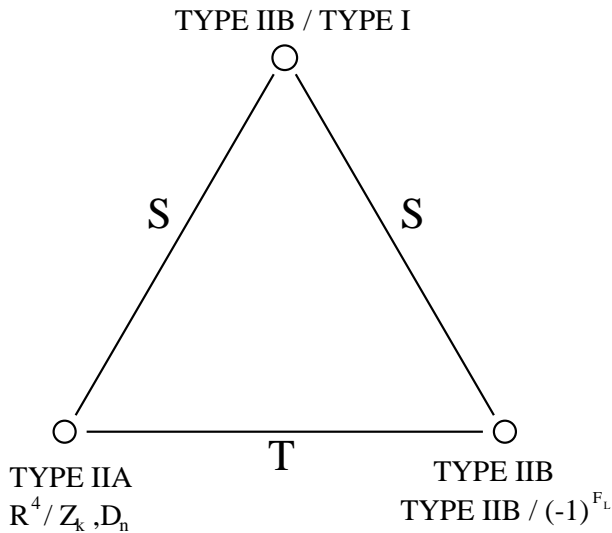}} \par}

\caption{A graph consisting of commuting T and S dualities. The T-duality between Type
IIA at an orbifold singularity and Type IIB with NS-branes can be represented
as two successive strong-weak coupling dualities. \label{chain}}
\end{figure}

\item T duality. \( ON^{0} \) is also connected to the T-dual of a Type IIA \( D_{n} \)
singularity. The proposal is that a T-duality along one of the directions of
a Type IIA \( D_{n} \) singularity is the Type IIB orbifold \( R^{4}/(-1)^{F_{L}}R \)
in the presence of \( n-1 \) physical NS-branes. The gauge symmetry is in both
cases \( SO(2n) \). The \( n \) twisted states of the \( D_{n} \) Type IIA
singularity, which are \( (1,1) \) six-dimensional vector multiplets, are mapped
to the \( n-1 \) multiplets living on the NS-branes in Type IIB plus the extra
multiplet living on the plane \( ON^{0} \). This proposal can be motivated
by representing a T-duality as a combination of two strong-weak coupling dualities,
as depicted in figure \ref{chain}. A background with a \( D_{n} \) singularity
can be described as a particular decompactification limit of Type IIA on a singular
\( K3 \), which is dual to the Type I\( ' \) string theory on \( T^{4} \).
In the relevant decompactification limit, the dual theory can be represented
as Type IIB on \( R^{4}/\Omega R \). In this dual description, the non-perturbative
enhanced \( SO(2n) \) gauge symmetry of Type IIA is described by \( n \) physical
D5-branes near an \( O5^{-} \) orientifold plane. Another S-duality now brings
us back to the Type IIB orbifold \( R^{4}/(-1)^{F_{L}}R \) in the presence
of NS-branes. Similar arguments apply for Type IIB singularities and Type IIA
\( ON \) planes and NS branes. 
\end{itemize}
The importance of \( ON^{0} \) both as the strong coupling of \( O5 \) plane
and as an ingredient in the T-dual description of \( D_{n} \) singularities
is manifest in figure \ref{chain}. As strong coupling of an \( O5^{0} \) plane,
\( ON^{0} \) plays an important role in the study of mirror symmetry in three
dimensional \( N=4 \) gauge theories. As ingredient in the T-dual description
of \( D_{n} \) singularities, it plays an important role in better understanding
several brane configurations, varying from Brane Box Models to six-dimensional
theories.

But, before constructing explicit examples, we need to explain what is the behavior
of D-branes in the presence of an \( ON^{0} \) plane.

\subsection{D-branes in the presence of an \protect\( ON^{0}\protect \) Plane\label{pro}}

Consider a Dp-brane (\( p\leq 6 \)) transverse to the \( ON^{0} \) plane,
or the Type II orbifold \( R^{4}/(-1)^{F_{L}}R \). Let us take, for definiteness,
the \( R \) inversion acting on the coordinates \( 6,7,8,9 \).
\begin{figure}
{\par\centering \resizebox*{10cm}{!}{\includegraphics{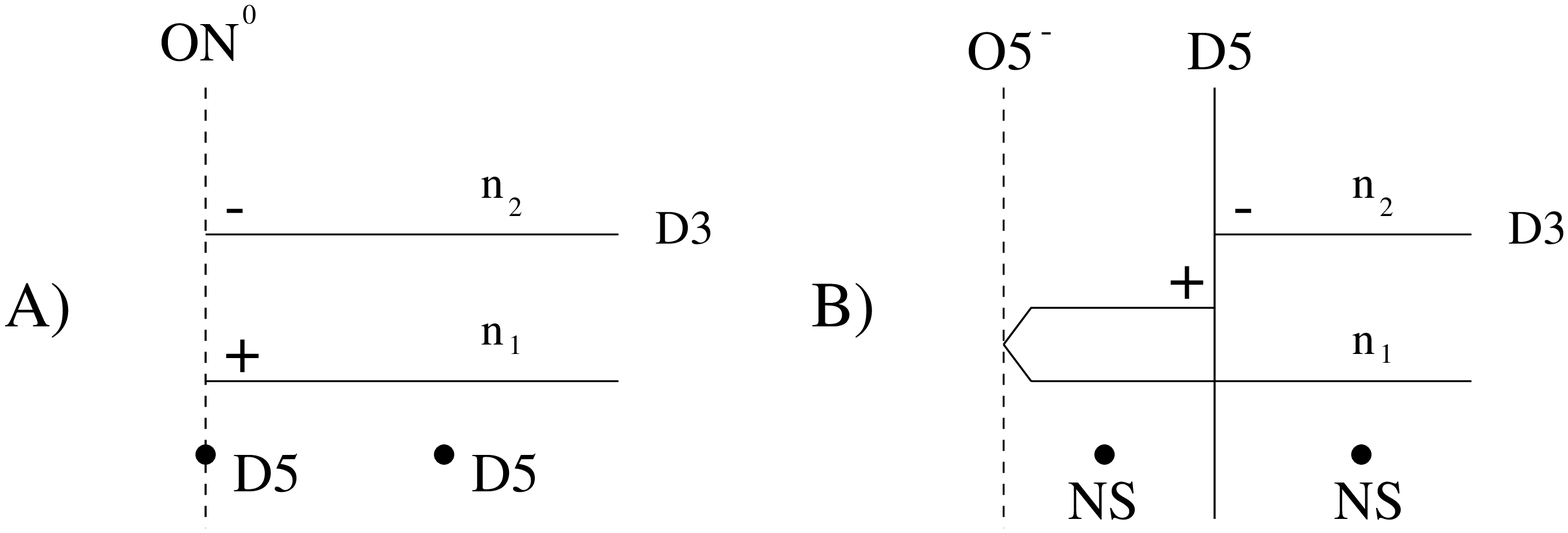}} \par}

\caption{Two useful ways of thinking of \protect\( ON^{0}\protect \) planes. In the
S-dual description (figure B) the different signs for the charges are explained
by considering branes ending on the D5 brane from the right or from the left.
We may also put extra D5-branes (and NS-branes in the S-dual configuration)
for future reference. In figure A, we put two kinds of D5-branes (represented
as dots). One of them is bounded to the \protect\( ON^{0}\protect \) plane;
in the S-dual picture (B) it has the interpretation of a NS-branes which lives
in between \protect\( O5^{-}\protect \) plane and the D5-brane.\label{LAST}}
\end{figure}

The crucial point is that the twisted sector of the \( (-1)^{F_{L}}R \) orbifold
has the same field content as a NS-brane and, therefore, a D-brane can end on
it. This means that the D-brane is a source for the fields living at the fixed
plane\footnote{
Notice that this is different from the standard configuration in which a D-brane
crosses an orientifold plane. Since the orientifold plane does not carry any
field on its world-volume, the D-brane is not really ending on it. 
}. As noticed in \cite{senS}, D-branes ending on the orbifold fixed plane may
have both positive and negative charge under the twisted fields; moreover, configurations
of parallel D-branes with different charges are supersymmetric \cite{senS,kapu}.
The reader should not confuse these negatively charged D-branes with anti D-branes.
Anti D-branes have a negative charge under the ten-dimensional RR-forms. In
this paper we always consider D-branes. Our branes are charged both under the
ten-dimensional RR-form (with positive charge) and under the six-dimensional
twisted sector form (with positive or negative charge).

The existence of different charges and the fact that supersymmetry is preserved
can be easily understood in the case \( p=3 \) going to the S-dual configuration
\cite{senS}. The result for general \( p \) follows from T duality. Consider
figure \ref{LAST}. In the S-dual picture, the orbifold plane is represented
by an orientifold \( O5^{-} \) plane and a physical D5-brane which can be moved
away from the orientifold point. The twisted sector of the orbifold point are
mapped to the fields on the D5-brane. D3-branes ending on the orbifold plane
are now ending on the D5-brane. Their charge under the D5-fields has a different
sign according to whether they end to the left or to the right of the D5-brane.
We can identify the positively charged branes ending on the orbifold plane with
the D3-branes ending on the D5-brane from the right, while the negatively charged
branes with D3-branes coming from the right infinity, going straight to the
orientifold, coming back and ending on the D5-brane from the left. This configuration
is manifestly supersymmetric.
\begin{figure}
{\par\centering \resizebox*{10cm}{!}{\includegraphics{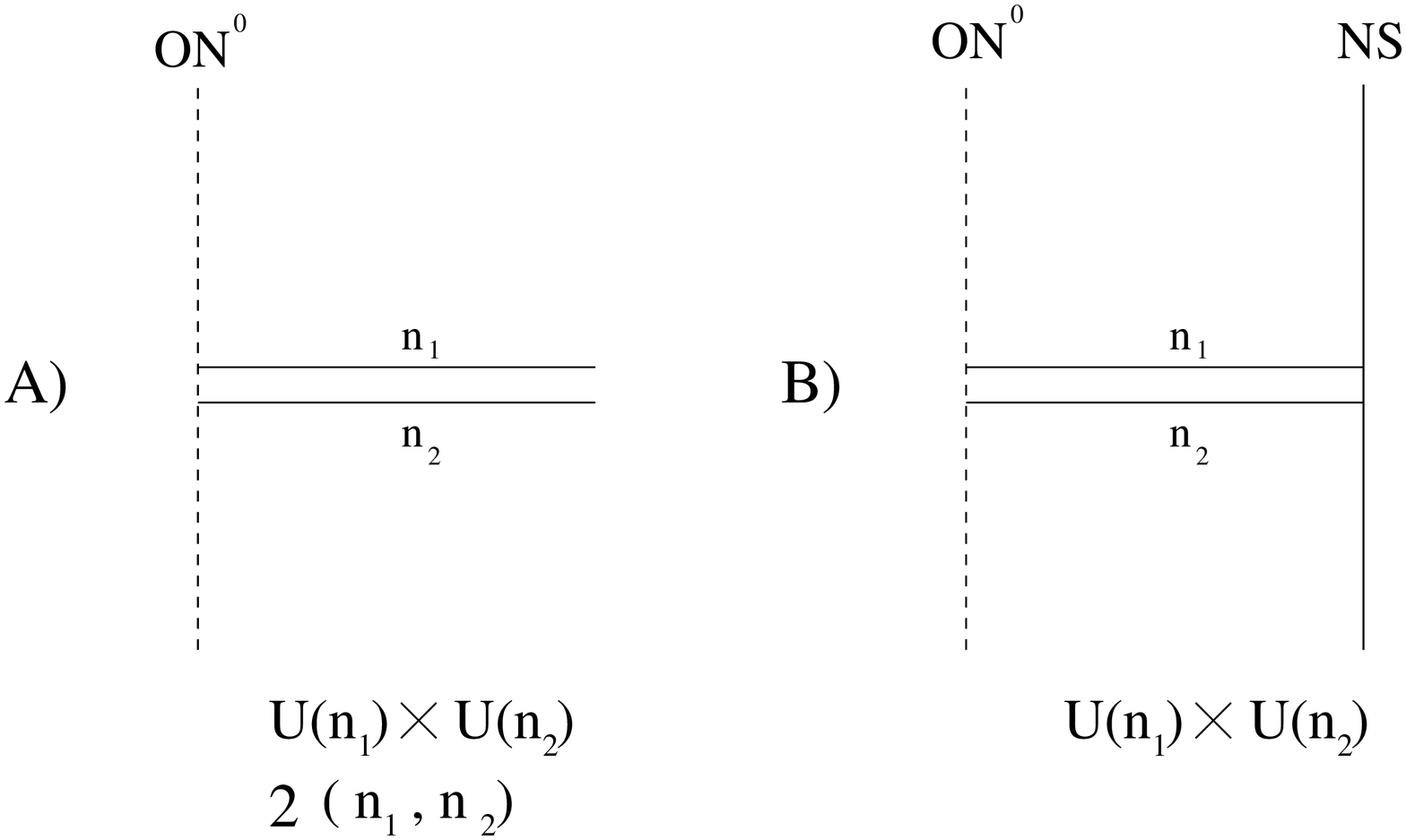}} \par}

\caption{Two sets of D-branes (\protect\( n_{1}\protect \) with positive charge and
\protect\( n_{2}\protect \) with negative charge) ending on \protect\( ON^{0}\protect \)
plane. The resulting gauge theory and matter fields are indicated below the
figure. In figure A there are two hypermultiplets in the bi-fundamental representation
of the two gauge groups. These hypermultiplets parametrize fluctuations transverse
to the orbifold plane. In figure B, the two hypermultiplet is projected out
by the presence of the NS brane.\label{rule}}
\end{figure}

\( n \) Dp-branes ending on the fixed plane, all of them with the same charge,
have a world-volume theory with eight supersymmetries consisting of a \( U(n) \)
gauge theory. There are no matter fields, since the hyper-multiplets corresponding
to the fluctuations transverse to the orbifold plane are projected out by the
action of \( R \). The rules for projecting the open string spectrum in the
case with both positive and negative charges were found in \cite{senS,kapu}
by using a boundary-state method. The rule is that, if there are \( n_{1} \)
Dp-branes with positive charge and \( n_{2} \) Dp-branes with negative charge
ending on the fixed plane, \( (-1)^{F_{L}}R \) acts on the open string Chan-Paton
factor as the conjugation by a diagonal matrix with \( n_{1} \) entries equal
to \( +1 \) and \( n_{2} \) equal to \( -1 \). This means that the gauge
fields (and their superpartners) coming from open strings connecting D-branes
with different charge are projected out and the gauge group is \( U(n_{1})\times U(n_{2}) \).
The hyper-multiplets corresponding to open strings with both ends on D-branes
with the same charge are projected out being odd under \( R \), but the ones
associated with \textit{mixed} open strings have an extra minus sign and survive
giving matter fields in the bi-fundamental \( (n_{1},n_{2}) \) representation
(figure \ref{rule}). The same result can be obtained by looking at the S-dual
configuration described in figure \ref{LAST}.
\begin{figure}
{\par\centering \resizebox*{17cm}{!}{\includegraphics{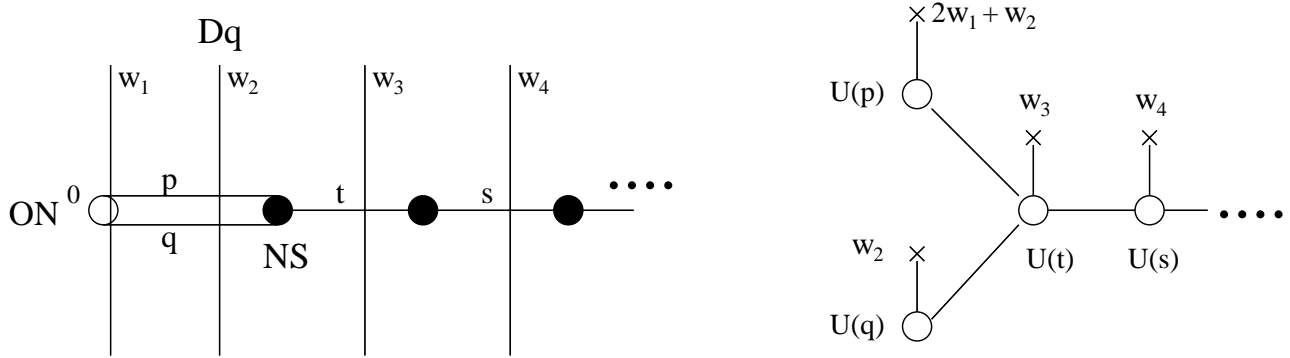}} \par}

\caption{The gauge theory can be read from the quiver diagram on the right. Nodes represent
the gauge group factors, links represent bi-fundamental matter fields and external
lines represent fields in the fundamental representation of the corresponding
gauge group.\label{example1}}
\end{figure}

In this paper we will need configurations with more NS-branes and with other
Dq-branes. They are depicted in figure \ref{example1} A. When there are both
Dp and Dq-brane, it is better to use pictorial conventions in which Dp-branes
are horizontal lines, NS-branes are points and Dq-branes or \( Oq \) planes
are vertical lines. Here \( q \) and the directions in which the Dq-brane extend
are chosen in such a way that preserves supersymmetry in the presence of Dp-branes;
the number of directions of type D-N in the open strings connecting Dp and Dq-branes
must be 4. Dp-Dq configurations considered in this paper are the D3-D5 systems
relevant for mirror symmetry in three dimensions, the D5-D7 systems for Brane
Box Models, and the D6-D8 systems for the construction of six-dimensional theories.\footnote{
See section \ref{section: oointer} for the explicit directions for the Dp and
Dq branes.
} The Dq-branes serve for introducing matter fields in the fundamental representation
of the various gauge groups. The world-volume theory obtained in this way can
be conveniently encoded in a quiver diagram, where nodes represent the gauge
group factors, links represent bi-fundamental matter fields and external lines
represent fields in the fundamental representation of the corresponding gauge
group; the quiver is depicted in figure \ref{example1} B. 

The curious splitting of the Dq-branes near \( ON^{0} \) requires some explanation.
The \( w_{1} \) Dq-branes in figure \ref{example1} A are \textit{bounded}
to the \( ON^{0} \) plane. The existence of bounded branes can be easily understood
in the particular case of \( p=3 \) and \( q=5 \); in the S-dual picture of
figure \ref{LAST} B, they are described by the NS-branes which live in between
\( O5^{-} \) and the D5-brane. It is obvious from figure \ref{LAST} that they
only contribute flavors to the \( U(p) \) gauge group. The \( w_{2} \) Dq-branes
are instead mapped to NS-branes which live to the right of the D5-branes; they
contribute flavors to both the \( U(p) \) and \( U(q) \) gauge groups. 

The global symmetry for a generic gauge factor is \( U(w_{i}) \). The gauge
theory has a \( U(2w_{1})\times U(w_{2}) \) symmetry for the factors associated
with branes near \( ON^{0} \). Notice that only a subgroup \( USp(2w_{1})\times U(w_{2}) \)
of this global symmetry is \emph{manifestly} realized in the brane picture.
Further arguments in favor of this symmetry will be discussed in section \ref{mirror}.

This kind of quiver theories naturally represent a small \( U(w) \) instanton
sitting on a \( D_{n} \) ALE space \cite{hm}; here \( w \) is the total number
of Dq-branes. The configuration with NS-branes is the T dual of the one considered
in \cite{hm}, as will be discussed in section \ref{T}. The form of the theory
is the same for every value of \( p \).

\subsection{Including Orientifold Planes\label{section: oointer}}

In this subsection we further complicate our life by introducing an orientifold
plane in addition to \( ON^{0} \). This configuration will be used in section
\ref{section:DD} for a particular class of mirror pairs in three dimensions,
and in section \ref{section: sixd} for discussing six-dimensional theories.
The reader not interested in these examples can skip this subsection.
\begin{figure}
{\par\centering \resizebox*{14cm}{!}{\includegraphics{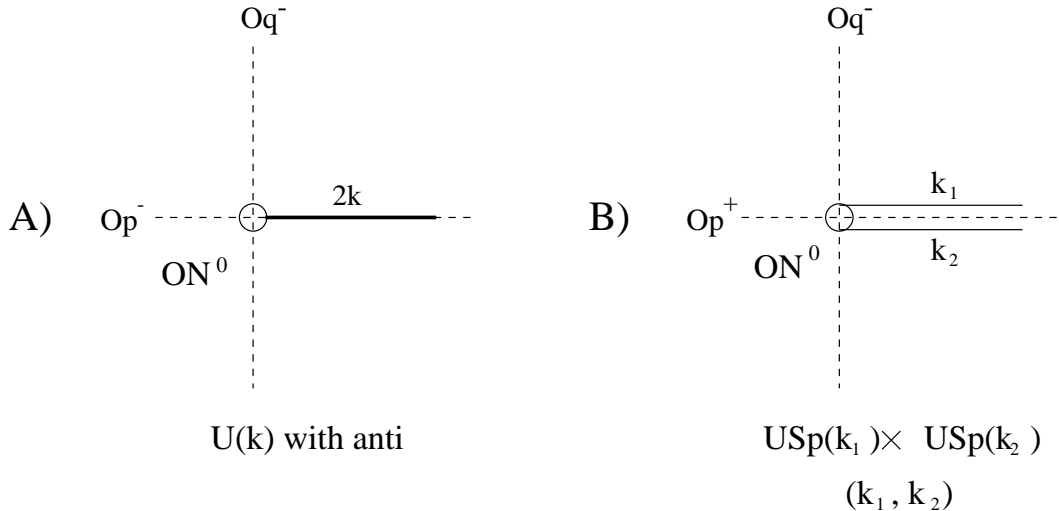}} \par}

\caption{In the case `without vector structure' (figure A), the states living on \protect\( ON^{0}\protect \)
and responsible for absorbing the charge of Dp-branes are projected out by \protect\( \Omega \protect \).
Therefore there is only one type of Dp-brane, living at the intersection of
\protect\( Oq^{-}\protect \) and \protect\( Op^{-}\protect \) planes. This
is the theory discussed in \cite{gp}. In the case `with vector structure' (figure
B), the surviving states living on \protect\( ON^{0}\protect \) allow and require
the existence of two kinds of Dp-branes. Each set of branes supports a
 \protect\( USp(k)\protect \)
group due to the existence of an \protect\( Op^{+}\protect \) plane.\label{projection}}
\end{figure}

We now discuss the behavior of Dp-branes near an \( ON^{0} \) plane when we
also introduce an \( Oq^{-} \) plane to the picture. 

In this paper, we consider \( ON^{0} \) as a perturbative orbifold projection;
in this, it differs very much from its natural partners, the NS-branes, which
are solitonic objects. The \( Oq^{-} \) and the \( ON^{0} \) plane combine
into a \( Z_{2}\times Z_{2} \) orbifold/orientifold projection, generated by
\( \Omega R_{q} \) and \( (-1)^{F_{L}}R_{6789} \), where \( R_{q} \) represents
a \( Z_{2} \) inversion in all the coordinates transverse to the \( Oq \)
plane. We are using notation in which the \( ON^{0} \) plane extends along
\( (012345) \) and the Dp-branes are stretched along \( (0,1,\ldots ,p-1) \)
and are possibly finite along \( x_{6} \). The \( Oq \) plane extends over
the coordinates \( (0,1,\ldots ,p-1) \) and \( (7,8,9) \). We will denote
\( R_{6789} \), in short, \( R \). 

We now discuss what kind of theory is realized on Dp-branes in this situation.
The reader not interested in technical details may want to skip the two following
paragraphs dealing with tadpoles, projections and all that, and look at figure
\ref{projection}. In the following we present two derivations for the massless
fields in this configuration.

\textbf{\emph{First}} \textbf{derivation}\textbf{\emph{\noun{.}}} Every generator
of the orbifold/orientifold projection acts on the Chan-Paton factors of the
D-branes according to the rules in \cite{gp,hm}. Under a T-duality in the directions
\( p,..,5 \) and \( 6 \) , the factor \( (-1)^{F_{L}} \) disappears \cite{senT},
and we recover a non-compact version of the original model in \cite{gp}. As
widely discussed in the literature \cite{hm,leigh,p,intblum1,intblum2}, there
are essentially two different consistent models. They differ in the perturbative
definition of the \( \Omega  \) projection on the closed string twisted states
and in the action of \( \Omega  \) on the Chan-Paton factors. Geometrically,
they are distinguished by the type of \( SO(2m) \) bundle that can be defined
on a space with a \( Z_{2} \) singularity \cite{leigh}. This \( SO(2m) \)
bundle is realized on the world-volume of the Dq-branes. A bundle defined on
a space with a \( Z_{2} \) singularity may admit vector structure, or not.
The original example in \cite{gp} does not admit vector structure. Modifications
that admit vector structure were constructed in \cite{gj, leigh,p,intblum1,intblum2}.
The difference between the two cases is encoded in the following relation between
the matrices that act on the Chan-Paton factors \cite{hm,p,intblum1,intblum2}:
\begin{equation}
\label{first}
\gamma _{\left[ (-1)^{F_{L}}R\right] }=\pm \gamma _{\Omega R_{q}}\gamma ^{T}_{\left[ (-1)^{F_{L}}R\right] }\gamma _{\Omega R_{q}}^{-1}\textrm{ }\, \, \, \, \, \, \, \, \, \, \, \, \, \, \, \, \textrm{with}/\textrm{without vector structure}
\end{equation}

Since we put an \( Oq^{-} \) plane, which by definition determines an \( SO(2m) \)
symmetry on the world-volume of Dq-branes, \( \gamma _{\Omega R_{q}} \) acts
as the identity matrix on Dq-branes \cite{gp}. The rule that the symmetry of
the matrix that projects Dp-branes is opposite to the one that projects Dq-branes
\cite{gp} is then enough to (almost) uniquely fix the projection matrices,
modulo an irrelevant change of basis. The result is that, in the case without
vector structure, the world-volume theory for \( 2N \) Dp-branes is \( U(N) \)
with antisymmetric matter fields, and, in the case with vector structure, is
\( USp(2N)\times USp(2N) \) with a bi-fundamental matter field. The sign in
equation (\ref{first}) is related to a global sign in the action of \( \Omega  \)
on the closed string twisted sector. The projection by \( \Omega R_{q} \) breaks
the Lorentz-invariance of the six-dimensional theory of the twisted states.
To avoid confusion, it is better to work in the particular case in which \( p=6 \),
in which this Lorentz invariance is not broken; this is the configuration of
D6 and D8 branes relevant for the study of six-dimensional theories. The result
for generic Dp and Dq-branes near \( ON^{0} \) follows from T-duality. The
twisted sector provides a tensor multiplet and a hyper-multiplet. In the case
without vector structure the tensor multiplet is projected out and the hypermultiplet
survives \cite{p}. The opposite happens in the case with vector structure.
Undoing the T-duality for discussing the case with generic \( p \), we discover
that, in the case without vector structure, the twisted state which may absorb
the charge of a Dp-brane is projected out by the \( \Omega  \) projection,
while in the case with vector structure it survives. The sign in equation (\ref{first})
is also related to the sign of the RR charge of the \( Op \) plane that is
induced in the theory. It is the standard \( Op^{-} \) plane in the case without
vector structure, but it is \( Op^{+} \) in the case with vector structure
\cite{p}. 

\textbf{\emph{Second derivation.}} In \cite{hzsix} we exploited a general method
for studying the consistent Type II and Type I models, living at \( Z_{n} \)
singularities, where differences between the models (presence or absence of
vector structure) were manifest in a dual brane description. Locally, the system
\( ON^{0}-Oq \) is indistinguishable from the generic \( Z_{2} \) orbifold/orientifold,
which was studied in \cite{hzsix}. In the next sections, we will add more NS-branes
and the methods in \cite{hzsix} will not be sufficient for describing the full
system, but, at that time, we should already know how to deal with \( ON^{0} \).
The methods in \cite{hzsix} allow to study the possible \( Z_{2} \) orbifold/orientifold
models in a T dual picture where the singularity is replaced by two NS-branes.
\begin{figure}
{\par\centering \resizebox*{12cm}{!}{\includegraphics{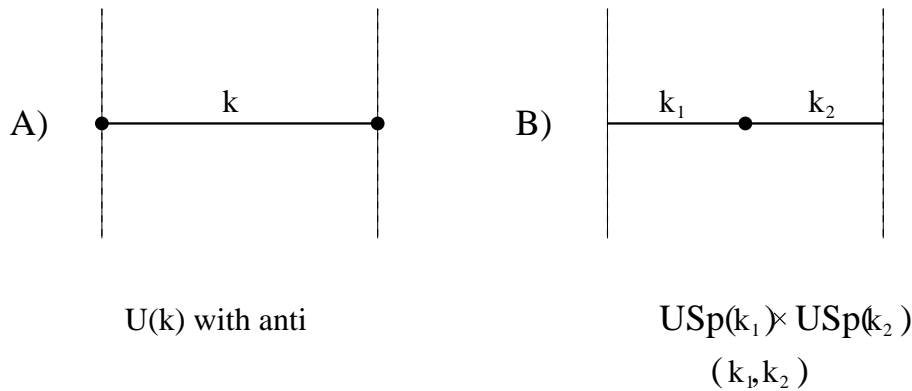}} \par}

\caption{\protect\( Z_{2}\protect \) orbifold/orientifold models, without (A) or with
vector structure (B). Details are exhaustively discussed in \cite{hzsix}. Lines
are Dq-branes and points are NS-branes. \label{zitwo}}
\end{figure}

To avoid confusion, we again specialize our discussion to the case with \( p=6 \);
a T-duality gives the result for generic \( p \). The two ways of putting two
NS-branes on a circle, respecting the \( Z_{2} \) symmetry, give two possible
models, depicted in figure \ref{zitwo}. The gauge group is, in the case without
vector structure, \( U(N) \) with antisymmetric matter, and, in the case with
vector structure, \( USp(N)\times USp(N) \) with a bi-fundamental. In the first
case, there are no tensor multiplets, while in the second case there is one.
Applying a T-duality for recovering the case with generic \( p \), we discover
that, in the case without vector structure, the twisted state which may absorb
the charge of a Dp-brane is projected out by the \( \Omega  \) projection,
while in the case with vector structure it survives. 

\vskip 0.3truecm 

The results of this long analysis are summarized in figure \ref{projection}.
The results are consistent with a naive extrapolation of the rules discussed
in section  \ref{pro} to the case with an orientifold. The crucial point of
the previous analysis is that the states associated with \( ON^{0} \), responsible
for the splitting of the branes, survive the orientifold projection only in
the case with vector structure. Therefore, we can have two kinds of Dp-branes,
and the typical D-quiver splitting only in the case with vector structure. In
the case without vector structure, we have only one kind of brane, which is
projected according to the original example in \cite{gp}. 
\begin{figure}
{\par\centering \resizebox*{17cm}{!}{\includegraphics{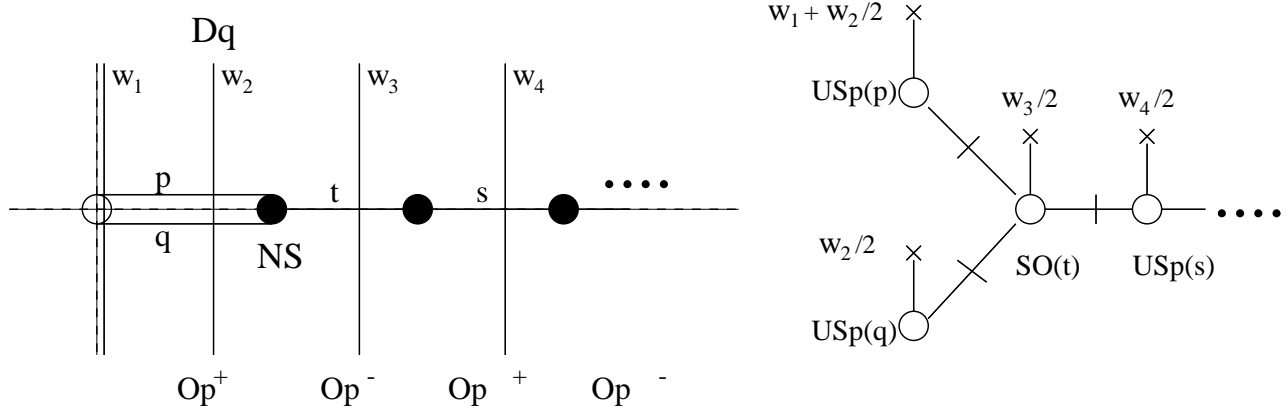}} \par}

\caption{The gauge theory can be read from the quiver diagram on the right. The bar
on the links means that they give rise to a half bi-fundamental.\label{example}}
\end{figure}

Consider now the addition of NS-branes. This gives the theory depicted in figure
\ref{example}. We face a lot of small subtleties. The following are the general
rules for determining the matter content: 

\begin{enumerate}
\item At every NS-brane the charge of the \( Op \) plane changes \cite{charge}.
Gauge groups associated to \( 2k \) Dp-branes near \( Op^{+} \) are \( USp(2k) \),
while those associated to \( k \) branes near \( Op^{-} \) are \( SO(k) \).
The only exception occurs when an \( Op^{-} \) crosses \( ON^{0} \); in this
case the gauge group associated to \( 2k \) Dp-branes is \( U(k) \). 
\item The bi-fundamental \( (p,q) \) fields appearing in figure \ref{projection},
associated to the branes near \( ON^{0} \), are absent when the branes end
on a NS-brane, since the latter freezes all fluctuations transverse to itself.
This is the general case, unless one considers the degenerating examples associated
to a singularity of type \( D_{2} \). 
\item There are only half bi-fundamentals for the gauge groups \( USp\times SO \).
This is due to the \( Op \) projection. This rule does not apply for gauge
groups \( USp\times U \) or \( SO\times U \). 
\item The Dq-branes can be partitioned among the NS-branes and give rise to matter
fields in the fundamental representation. \( w_{i} \) Dq-branes located where
an \( Op^{+} \) plane exists give rise to an \( SO(w_{i}) \) global symmetry,
while those located where an \( Op^{-} \) plane exists give rise to a \( USp(w_{i}) \)
global symmetry. This global symmetry teaches us that \( w_{i},(i\neq 1,2) \)
Dq-branes give rise to \( \frac{w_{i}}{2} \) flavors. Near \( ON^{0} \) (\( w_{1} \)
and \( w_{2} \) ) there is a further splitting for the Dq-branes. This is analogous
to the splitting encountered in section \ref{pro}. The global symmetry is \( SO(2w_{1})\times SO(w_{2}) \).
The field theory may get a bigger global symmetry \( SO(2w_{1}+w_{2})\times SO(w_{2}) \)
not seen by the branes. 
\end{enumerate}
This kind of quiver theories naturally represents small \( SO(w) \) instanton
on a \( D_{n} \) ALE space \cite{hm}. The configuration with NS-branes is
the T dual of the one considered in \cite{hm}, as will be discussed in section
 \ref{T}. The form of the theory is the same for every value of \( p \). In
the particular case \( p=6 \) the world-volume theory is uniquely specified
by the requirement of anomaly cancellation. This will be exhaustively discussed
in section  \ref{section: sixd}.

\section{T-Duality for \protect\( D_{n}\protect \) Singularities\label{T}}

In this Section, we discuss in more detail the T-dual of a \( D_{n} \) singularity.
It is well known that, in Type II string theory, a T-duality along one of the
directions of an ALE space of type \( A_{k} \) transforms it into a Type II
flat background with \( k+1 \) NS-branes \cite{vaoo}. As already anticipated,
the proposal is that the T-dual configuration for an ALE space of type \( D_{n} \)
is the Type II orbifold \( R^{4}/(-1)^{F_{L}}R \) in the presence of \( n-1 \)
physical NS-branes. 

In the \( A_{k} \) case, the \( k+1 \) twisted states of a \( Z_{k+1} \)
orbifold\footnote{
The twisted states are vector supermultiplets of a \( (1,1) \) six-dimensional
theory in Type IIA and \( (2,0) \) tensor supermultiplets in Type IIB. 
} are mapped to the world-volume fields of the \( k+1 \) NS-branes\footnote{
NS-branes indeed support vector multiplets in Type IIB and tensor multiplets
in Type IIA. 
}. In the \( D_{n} \) case, the NS-branes world-volume fields provide the dual
description of only \( n-1 \) twisted states. The dual of the remaining twisted
state comes from the fixed point of \( (-1)^{F_{L}}R \); the twisted sector
of \( (-1)^{F_{L}}R \), both in Type IIA and Type IIB, has indeed the same
massless field content as a NS-brane.
\begin{figure}
{\par\centering \resizebox*{17cm}{!}{\includegraphics{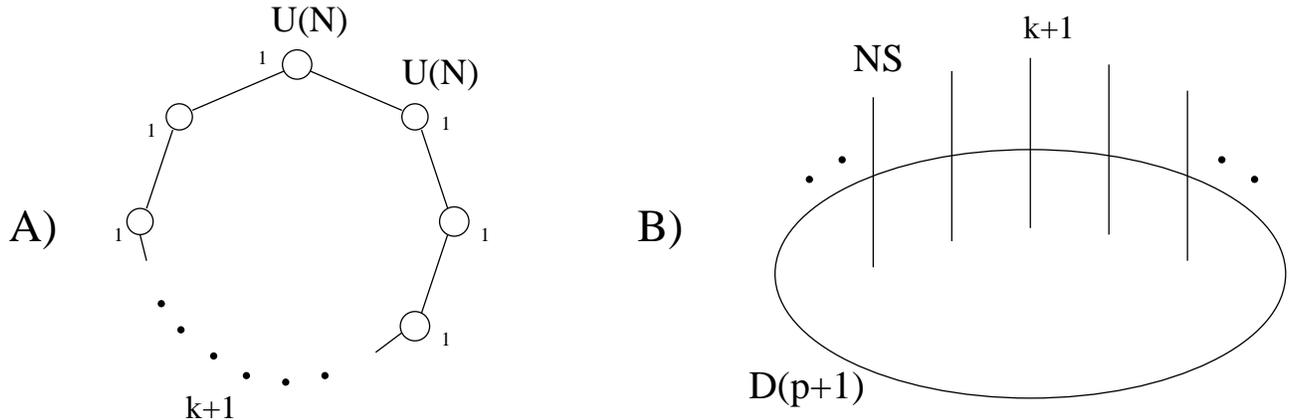}} \par}

\caption{In figure A, the extended Dynkin diagram for \protect\( A_{k}\protect \) is
depicted. The small numbers near each node are the Dynkin indices. Each node
is associated with a gauge group factor and each link with a hypermultiplet
in the bi-fundamental representation of the corresponding groups. The resulting
theory is \protect\( U(N)^{k+1}\protect \) with bi-fundamentals for neighboring
\protect\( U(N)\protect \) factors. The same theory can be obtained by considering
figure B, which describes D-branes wrapped around a circle and broken in between
\protect\( k+1\protect \) NS-branes.\label{braneA}}
\end{figure}

It is sometimes useful to replace the ALE space with a Taub-NUT geometry (also
called KK monopole). Globally the structure of an ALE space is \( R^{4} \),
that of a Taub-NUT space is \( R^{3}\times S^{1} \), which makes it more suitable
for performing T-dualities. The ALE space can be recovered by sending a parameter
(which roughly measures the radius of the circle) to infinity. The T-dual of
a \( D_{n} \) KK monopole is the \( R^{3}\times S^{1}/(-1)^{F_{L}}R \) orbifold
with \( n-2 \) NS-branes. Since there are two fixed points of \( (-1)^{F_{L}}R \)
along \( S^{1} \), only \( n-2 \) NS-branes are now required to match the
\( n \) twisted states. The T-dual of the \( D_{n} \) orbifold is obtained
by sending the radius of the dual \( S^{1} \) to zero.

The proposal was motivated before using the chain of dualities depicted in figure
\ref{chain}. Related arguments can be found in earlier literature \cite{senT,kut},
where it was noticed that the perturbative T dual of the singular \( K3 \)
manifold \( T^{4}/Z_{2} \) is \( T^{4}/(-1)^{F_{L}}Z_{2} \) and a description
in terms of NS-branes was proposed.

We can now ask what happens when we introduce D-brane probes in the picture.
There is an almost complete dictionary for determining the world-volume theory
of D-branes sitting at ALE orbifold singularities \cite{hm}. \( N \) Dp-branes
sitting at a singularity of the form \( R^{4}/\Gamma _{G} \), where \( G \)
is a simply-laced group, have a world-volume theory that is associated with
the extended Dynkin diagram for G, with a gauge factor for each node of the
diagram and a bi-fundamental matter field for each link \cite{hm}. The gauge
group is \( \prod U(n_{\mu }N) \), where \( n_{\mu } \) are the Dynkin indices
for the group G. The Higgs branch of these theories, which is the same for all
\( p \) and is not corrected by quantum effects, is the symmetric product of
\( N \) copies of the ALE space; this is the brane realization \cite{hm} of
the well known mathematical construction of the ALE spaces as hyperK\textbf{\( \ddot{a} \)}hler
quotients. 

We should be able to see that the world-volume theory of D-branes probe is preserved
by the previously discussed T-duality. The analysis of the \( A_{k} \) case
is straightforward. N Dp-branes (\( p\leq 6 \)) sitting at a \( Z_{k+1} \)
singularity have a world-volume theory \( U(N)^{k+1} \) with bi-fundamentals
for neighboring \( U(N) \) factors. Deforming the ALE space to a Taub-NUT and
performing a T-duality along the \( S^{1} \), we obtain a configuration of
\( N \) D(\( p+1 \))-branes in the presence of \( k+1 \) NS-branes, configuration
that was discussed in \cite{hw} and reproduces the same world-volume theory
(figure \ref{braneA}).
\begin{figure}
{\par\centering \resizebox*{17cm}{!}{\includegraphics{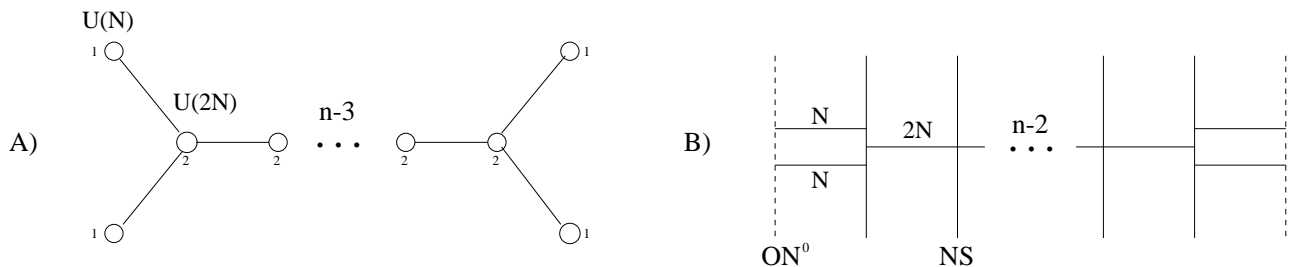}} \par}

\caption{In figure A, the extended Dynkin diagram for \protect\( D_{n}\protect \) is
depicted. The small numbers near each node are the Dynkin indices. Each node
is associated with a gauge group factor and each link to a hypermultiplet in
the bi-fundamental representation of the corresponding groups. The resulting
theory is \protect\( U(N)^{2}\times U(2N)^{n-3}\times U(N)^{2}\protect \) with
bi-fundamentals for neighboring factors. The same theory can be obtained by
considering figure B, which describes D-branes wrapped around a circle with
two fixed points, here depicted as a segment. \label{braneB}}
\end{figure}

That the T-duality process gives a consistent result also in the \( D_{n} \)
case can be now easily shown using the results in the previous sections. The
world-volume theory for \( 2N \) Dp-branes sitting at a \( D_{n} \) singularity
is \( U(N)^{2}\times U(2N)^{n-3}\times U(N)^{2} \) with bi-fundamentals associated
with the links of the \( D_{n} \) extended Dynkin diagram (figure \ref{braneB}).
After T-duality, we have a set of \( 2N \) D(\( p+1 \))-branes wrapped around
a circle with two fixed points under \( (-1)^{F_{L}}R \). We can picture the
projected circle as a segment. We also have \( n-2 \) NS-branes. Combining
the methods in \cite{hw} with those described above, it is straightforward
to check that the theory associated to this configuration of branes is the same
as that associated to the \( D_{n} \) extended Dynkin diagram. 

Configurations in which \( M \) Dq-branes are also present are interesting.
We are considering a situation in which the Dp-probe is sitting at a point in
the ALE space, \( q=p+4 \) and the Dq-brane is wrapped on the ALE space. This
configuration naturally describes \( N \) small \( U(M) \) instanton on an
ALE space; the Higgs branch of Dp-branes world-volume theories, which is the
same for all values of \( p \) and is not corrected by quantum effects, is
isomorphic to the \( N \) instantons moduli space \cite{hm}. The case of \( SO(2M) \)
instantons is easily obtained by considering an \( Oq \) plane. After a T-duality,
we obtain a configuration with D(\( p+1 \)), D(\( q-1 \)) and NS-branes. The
world-volume theory can be read using the results in sections \ref{pro} and
\ref{section: oointer} and agrees with the one discussed in \cite{hm}. The
representation of small instantons sitting at space-time singularities as systems
of D and NS branes was discussed in \cite{hzsix} for the case of the \( A_{k} \)
ALE spaces. The replacement of space-time singularities with dual smooth backgrounds
with NS-branes has a double purpose. First, it allows to have a better control
on parameters and moduli of the theory. Second, it provides a very simple and
intuitive classifications of the gauge bundle compatible with the singularity;
it naturally distinguishes between bundles with or without vector structure
and accounts for the breaking to a subgroup due to the singularity. The analysis
in \cite{hzsix} can be easily extended to the \( D_{n} \) ALE space using
the results in sections \ref{pro} and \ref{section: oointer}. \( U(M) \)
instanton theories are associated to the configurations in figure \ref{example1}
and \( SO(2M) \) instanton theories to the configurations in figure \ref{example};
both the figures must be made compact by including a second \( ON^{0} \) plane.
Once again, the disposition of the NS-branes on the segment accounts for the
allowed \( U(M) \) or \( SO(2M) \) bundles and for the breaking of the gauge
group to a subgroup. The extension of the construction in \cite{hzsix} to the
\( D_{n} \) series is one of the achievements of this paper, but we do not
indulge more on this here, since section  \ref{section: sixd} will be devoted
to the subject. The small \( U(M) \) or \( SO(2M) \) instanton at \( D_{n} \)
singularities will also appear in the discussion about mirror symmetry.

\section{Mirror Symmetry\label{mirror}}

\begin{figure}
{\par\centering \resizebox*{17cm}{!}{\includegraphics{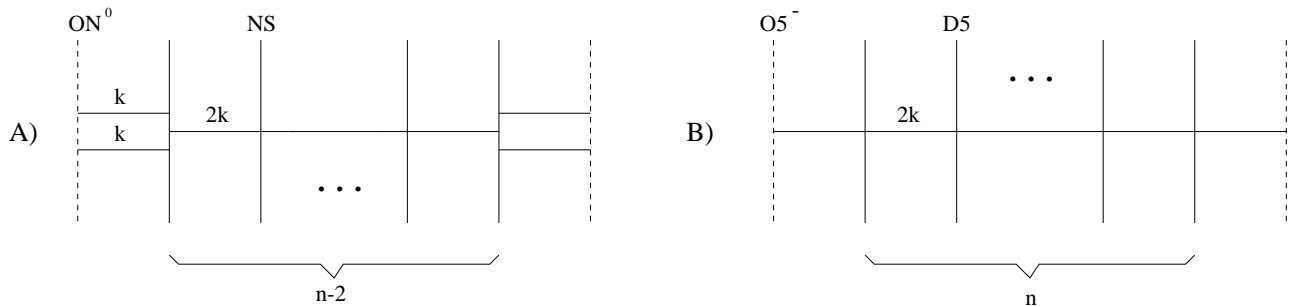}} \par}

\caption{Brane configuration for \protect\( USp(2k)\protect \) with an antisymmetric
and \protect\( n\protect \) flavors (B) and its mirror (A). \label{braneC}}
\end{figure}

The \( ON^{0} \) plane was used in \cite{kapu} to explain three-dimensional
mirror symmetry for \( USp(k) \) gauge group coupled to a hypermultiplet in
the second rank antisymmetric tensor representation and \( n \) flavors in
the fundamental representation. The candidate mirror theory, associated with
the \( D_{n} \) extended Dynkin diagram, was guessed in \cite{deboer} and
demonstrated in \cite{pz}, using M-theory. However, a description in terms
of the construction in \cite{hw} was still lacking and, with that, something
in our knowledge of the dictionary for translating general gauge theories in
terms of brane models was missing. The understanding of the strong coupling
of \( O5^{0} \) planes clearly closes this gap \cite{senS,kapu}. 

A \( D_{n} \) quiver three-dimensional \( N=4 \) gauge theory can be realized
as in figure \ref{braneC} A, as extensively discussed in section  \ref{T}.
It is simple, using the rules in \cite{hw}, to check that the mirror theory
is indeed \( USp(2k) \) with an antisymmetric and \( n \) flavors \cite{kapu}.
An S-duality transforms NS-branes into D5-branes and the \( ON^{0} \) plane
into a combination of an \( O5^{-} \) plane and a physical D5-brane. As a result,
an S-duality transforms the system in figure \ref{braneC} A into a set of D3-branes
stretched between two \( O5^{-} \) planes in the presence of \( n \) D5-branes,
as depicted in figure \ref{braneC} B. This is a fairly standard configuration
which realizes the above mentioned \( USp(2k) \) theory. For more details,
the reader is referred to \cite{kapu}. We will shortly show many examples that
should make clear this type of construction.

\subsection{Examples\label{examples} }

\begin{figure}
{\par\centering \includegraphics{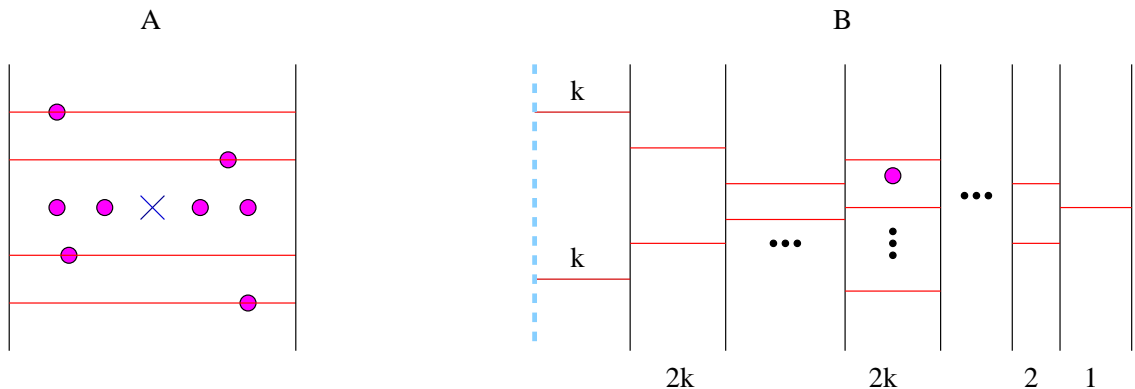} \par}

\caption{Brane realization for \protect\( USp(2k)\protect \) gauge theory with \protect\( n\protect \)
flavors in the fundamental representation. Figure A shows the coulomb branch
of the theory. This figure is drawn on the double cover of the projected space.
The \protect\( X\protect \) symbol in the middle denotes an \protect\( O5^{-}\protect \)
plane. The circles denote D5 branes. The vertical lines represent NS brane and
its image, the horizontal lines are D3 branes. Figure B, drawn only on the physical
plane, shows the Higgs branch of this theory, or alternatively the Coulomb branch
of the mirror. The branes in the figure are the S-dual of the branes in figure
A, with the same notation for the branes. The numbers below or next to a set
of horizontal lines denote the number of D3 branes in between two NS branes.\label{spwithflavors}}
\end{figure}

The examples that we explicitly discuss are: 

\begin{itemize}
\item Mirror of \( USp(2k) \) with \( n \) flavors in the fundamental representation. 
\item Mirror of \( USp(2k_{1})\times USp(2k_{2}) \) with a hypermultiplet in the
bi-fundamental representation, \( n_{1} \) flavors in the fundamental representation
of the first group and \( n_{2} \) flavors in the fundamental representation
of the second group. 
\item Mirror of \( U(2k) \) with one or two matter fields in the two-rank antisymmetric
tensor representation and \( n \) flavors in the fundamental representation. 
\end{itemize}
Many other examples and generalizations are also outlined.

\subsubsection{Mirror of \protect\( USp(2k)\protect \) with \protect\( n\protect \) Flavors\label{SP}}

The mirror theory was found in \cite{horivafa} using different methods; this
serves as a check to our method. Later we will study other examples for which
there are no known mirrors in the literature. This demonstrates the power of
this approach. 

\begin{figure}
{\par\centering \includegraphics{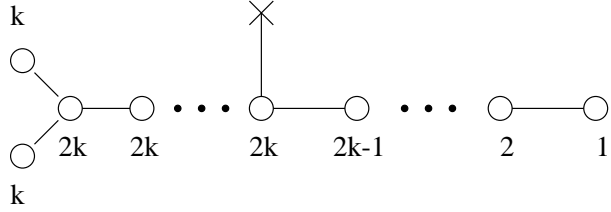} \par}

\caption{Dynkin diagram for \protect\( USp(2k)\protect \) with \protect\( n\protect \)
flavors. The numbers next to nodes denote the rank of the gauge factor in the
gauge group. Lines denote bi-fundamental hypermultiplets. A line ending on an
\protect\( X\protect \) denotes a hypermultiplet which is charged with respect
to the corresponding node it connects. Later this will be called an external
line. \label{spdynkin}}
\end{figure}

The procedure for calculating the mirror theory follows standard steps using
the rules in \cite{hw}. To move to the Higgs branch of the original theory,
one goes to the origin of moduli space, namely to the region in which the D5
branes touch the D3 branes. Once they touch, the D3 branes can split. Figure
\ref{spwithflavors} B shows the maximal splitting for the D3 branes. The splitting
is done taking into account that S-configurations are not supersymmetric. As
a result the NS brane, when located to the right of all branes has D3 brane
tails connecting it to \( 2k \) D5 branes. These tails do not represent massless
modes in the field theory. Using the observation that the position of the NS
brane is not a relevant parameter for the low energy field theory, we move the
NS brane to the left. Whenever it crosses a D5 brane, a D3 brane is removed.
After passing \( 2k \) D5 branes the NS brane is located as in figure B. Note
that in figure B the NS-brane has been denoted as a circle. This before denoted
a D5 brane; we have indeed performed an S-duality. At this point we learn that
the gauge group in the segment where the circle is located gets an extra hypermultiplet.
The final gauge group can be encoded into the quiver diagram depicted in figure
\ref{spdynkin}. The resulting gauge group is 
\[
U(k)^{2}\times U(2k)^{n-2k-1}\times \prod _{i=1}^{2k-1}U(i),\]
 with matter as in the figure. Note that the first \( U(2k) \) factor has an
additional hypermultiplet which is only charged under this group.

\subsubsection{Mirror of \protect\( USp(2k_{1})\times USp(2k_{2})\protect \)}

\begin{figure}
{\par\centering \resizebox*{14cm}{!}{\includegraphics{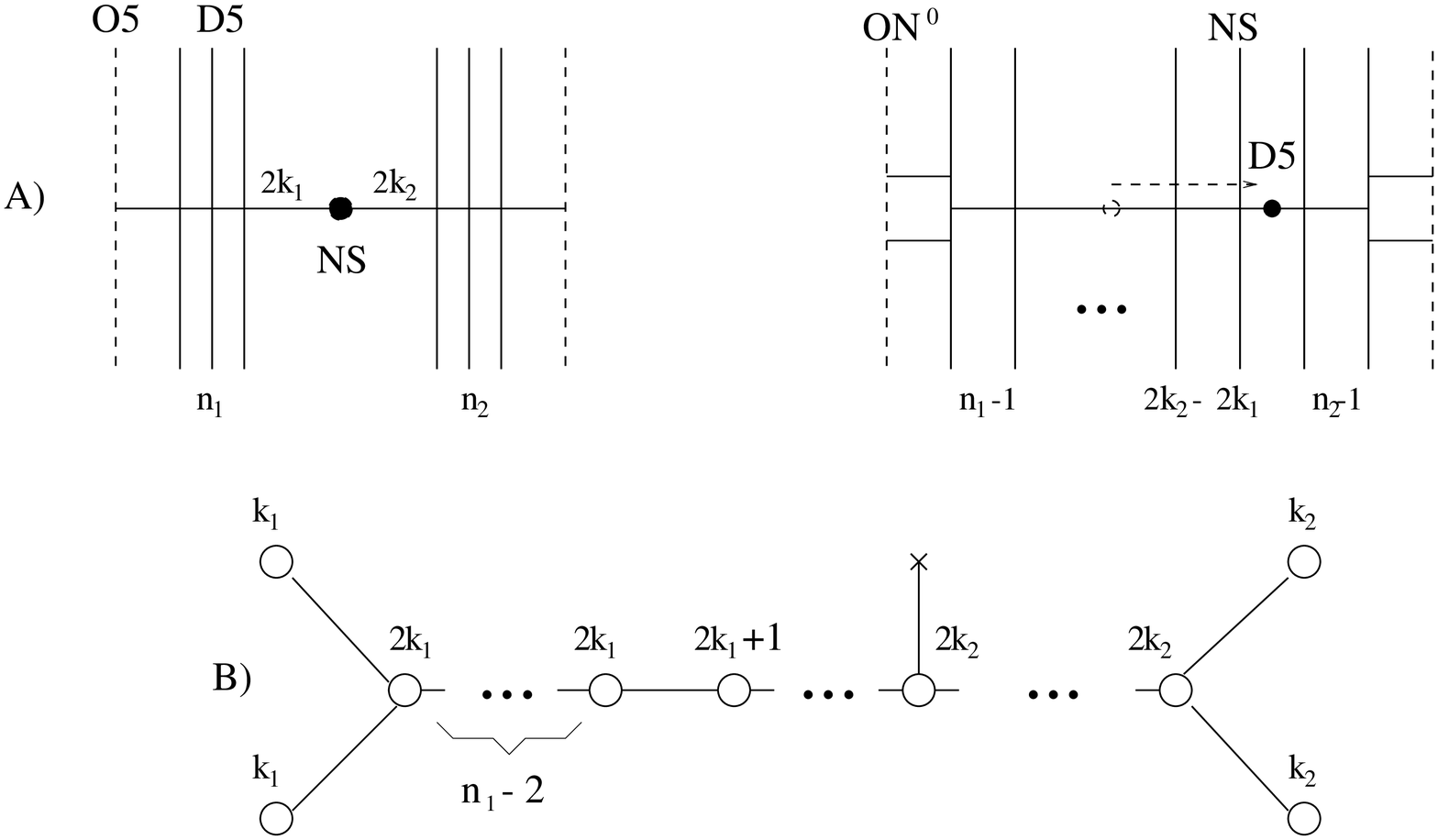}} \par}

\caption{The brane configuration for \protect\( USp(2k_{1})\times USp(2k_{2})\protect \)
with \protect\( n_{1}\protect \) flavors in the fundamental representation
of the first group, \protect\( n_{2}\protect \) flavors in the fundamental
representation of the second group and a bi-fundamental is shown in figure A
together with its mirror. The quiver diagram corresponding to the mirror theory
is shown in figure B.\label{double}}
\end{figure}

The next example is the mirror of \( USp(2k_{1})\times USp(2k_{2}) \) with
\( n_{1} \) flavors in the fundamental representation of the first group, \( n_{2} \)
flavors in the fundamental representation of the second group and a bi-fundamental.
The brane configuration is obtained by considering two \( O5^{-} \) planes;
it is depicted, with the S-dual configuration, in figure \ref{double} A. Notice
that we now switched to different notations, which are more suitable for giving
a synthetic description of compact models; to avoid confusion we explicitly
indicated the type of branes in the picture. The quiver diagram for the mirror
theory can be easily computed step-by-step using the rules in \cite{hw} and
it is shown in figure \ref{double} B. The particular case \( k_{1}=k_{2} \)
is almost trivial to compute; the reader simply needs to exchange NS- and D5-branes
and \( ON^{0} \) and \( O5^{0} \) planes. The case \( k_{1}\neq k_{2} \)
requires also some attention to non-supersymmetric configurations and to move
some of the D5-branes; the needed steps are very similar to the ones discussed
in the previous example. The resulting gauge group is, (assuming \( k_{1}>k_{2} \)
and \( n_{1}+2k_{2}>2k_{1}+1 \)),

\[
U(k_{1})^{2}\times U(2k_{2})^{n_{1}-2k_{1}+2k_{2}-1}\times \left[ \prod ^{2k_{1}-1}_{i=2k_{2}+1}U(i)\right] \times U(2k_{2})^{n_{2}-1}\times U(k_{2})^{2}.\]
 Special attention is required for the cases \( n_{1}+2k_{2}-2k_{1}=0,1 \).
Without loss of generality, we consider from now on \( k_{1}=k_{2} \). For
\( n_{1}=0,1 \), one should analyze the global symmetry of the three dimensional
theory, interpreted as the world volume theory which lives on the NS brane.
For generic \( n_{1} \), the NS brane supports the usual (1,1) supersymmetric
six dimensional \( U(1) \) multiplet. This is even true for the case \( n_{1}=1. \)
For \( n_{1}=0, \) we have enhanced symmetry which can be seen before making
the mirror operation. The NS brane has a modulus which measures the distance
from the \( O5^{-} \) plane. When this distance goes to zero a W-boson, given
by D-string stretching between the NS brane and its image under the \( O5^{-} \)
plane, becomes massless. The resulting group on the world volume of the NS brane
is enhanced to \( USp(2) \). More generally, \( n \) NS branes next to an
\( O5^{-} \) plane have an enhanced \( USp(2n) \) gauge theory on their world
volume. It should be noted that Poincar\'{ e} invariance in six dimensions
no longer holds since the \( O5^{-} \) plane breaks it explicitly. The world
volume theory of the NS brane is then a six dimensional theory with a point
like singularity at three of its coordinates. This breaks half of the supersymmetries
in addition to the explicit breaking of six dimensional Poincar\'{ e} invariance.

The case \( n_{1}=0 \) becomes even more special when we consider the configuration
after making the mirror transformation. The crucial point is that in the S-dual
picture the orientifold plane and a D5-brane combine into an \( ON^{0} \) plane.
For \( n_{1}=0 \), the NS-brane, which in the original configuration lives
between the orientifold and the D5-brane whose fate is to combine with it, becomes,
after S-duality, a D5-brane that is bounded to \( ON^{0} \). We already discussed
this kind of configurations in section  \ref{pro}. The enhanced symmetry on
the world volume theory of the D5 brane remains, as before the mirror transformation,
a \( USp(2) \) gauge theory. 

\begin{figure}
{\par\centering \resizebox*{16cm}{!}{\includegraphics{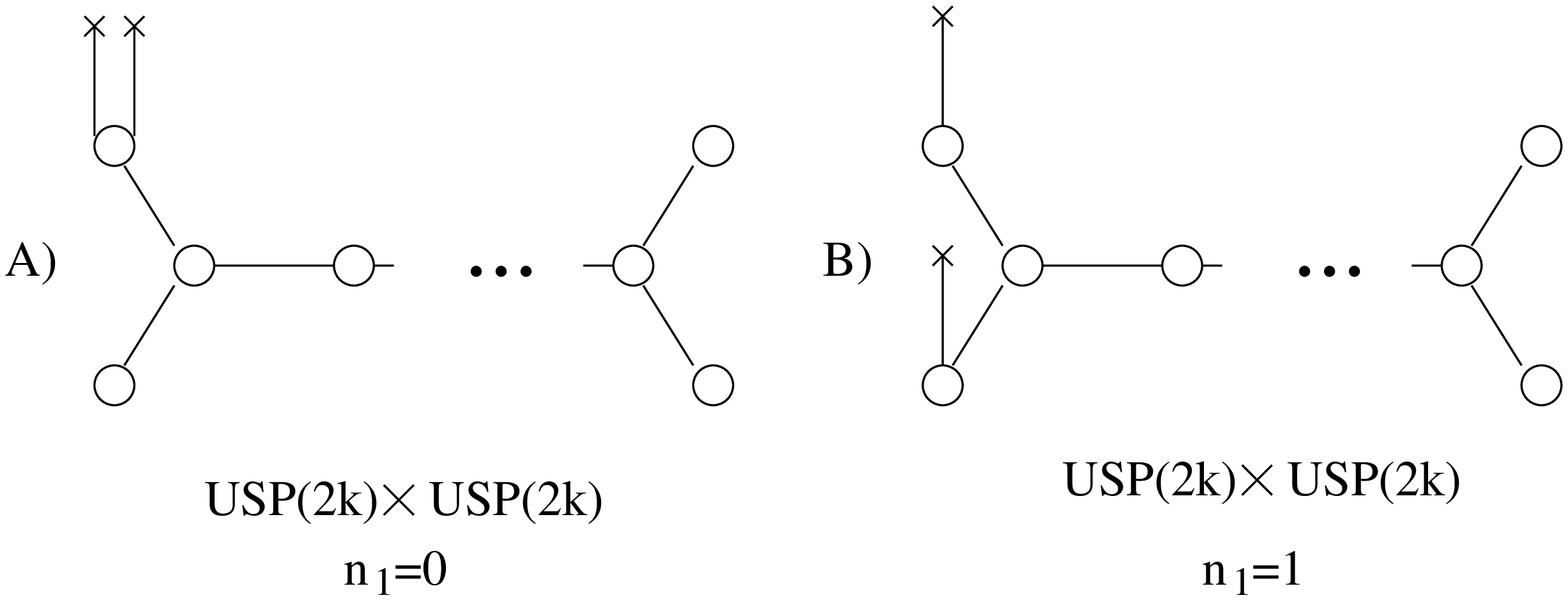}} \par}

\caption{Quiver diagrams of type D with some external lines. The mirror theories are
indicated below each graph. \label{cechs}}
\end{figure}

Combining this with the symmetry on the world volume of the D5-branes, we get
that the global symmetry of the problem for \( n_{1}=0 \) is \( USp(2)\times SO(2n_{2}) \).
For \( n_{1}\neq 0 \) the global symmetry is \( U(1)\times SO(2n_{1})\times SO(2n_{2}) \).
The knowledge of the global symmetry helps in deriving the quiver diagram corresponding
to the mirror theories. The diagrams for \( n_{1}=0,1 \) are drawn in figure
\ref{cechs} A and B.

\subsubsection{\protect\( SU(N_{c})\protect \) with Antisymmetric matter and \protect\( N_{f}\protect \)
Flavors.}

We have seen that there are examples with no external lines, corresponding to
figure \ref{double}, and with two external lines (figure \ref{cechs}). The
natural question to ask is how to get a quiver with only one external line.
The naive answer will be to get half of the case in figure \ref{cechs} B. Indeed
one needs to put a stuck NS brane on the \( O5 \) plane. This leads to a quiver
with one external line as required. What is the field theory of this configuration
of branes? Well, the answer is fairly standard. A stuck NS-brane gives a hypermultiplet
in an anti-symmetric tensor representation. On the other hand, the gauge theory
on the D3 brane still remains a \( USp \) gauge theory since it is projected
on the other side of the interval. So we get a theory with the same matter content
as for the case with no external lines as in figure \ref{double}. It is a known
phenomenon that the presence of a single NS brane does not change the matter
content but may make one of the parameters in the theory manifest as a deformation
of the brane configuration. 

\begin{figure}
{\par\centering \resizebox*{13cm}{!}{\includegraphics{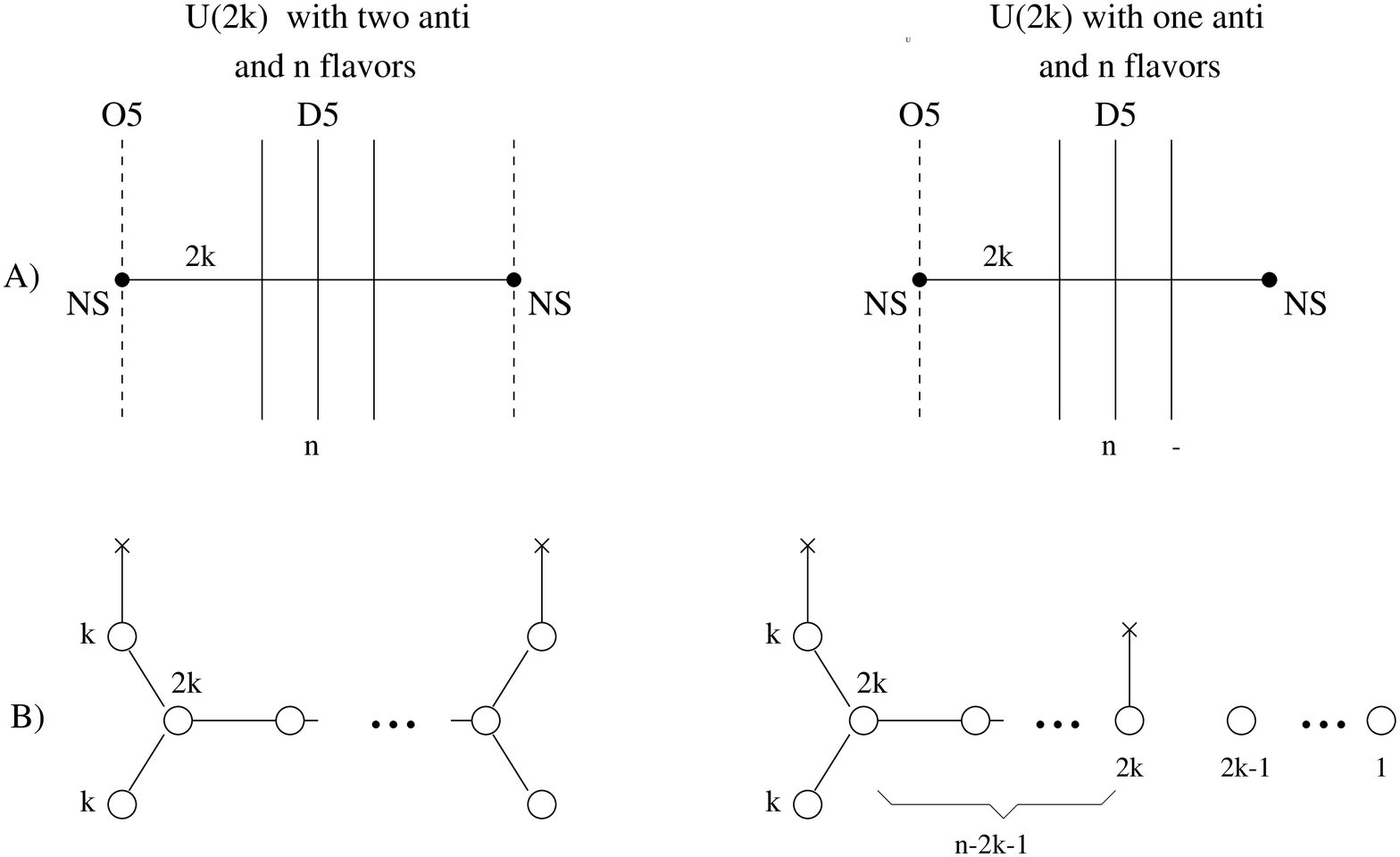}} \par}

\caption{Figure A: brane realization for \protect\( SU(N_{c})\protect \) gauge theory
with one or two flavors in the antisymmetric representation and \protect\( N_{f}\protect \)
flavors in the fundamental representation. Figure B: the quiver diagrams for
the mirror theories.\label{anti2}}
\end{figure}

The brane configurations for \( SU(2k) \) field theories with antisymmetric
matter are given in figure \ref{anti2} A. The steps for computing the mirror
theories closely parallel those considered in the previous examples. The computation
in the compact case is almost trivial; in the non-compact case, one needs to
pay attention to non-supersymmetric configurations and move branes around, closely
paralleling what is done in section \ref{SP}. The results for the mirror theories
can be conveniently encoded in the quiver diagrams shown in figure \ref{anti2}
B. There is no enhanced symmetry associated to the stuck NS branes.

All the considered compact models are associated to a \( D_{n} \) extended
Dynkin diagram, decorated with one or two external lines. A convenient classification
of these kind of theories is given by looking at the global symmetry; we discussed
examples with global symmetry that is a subgroup of \( USp(2)\times SO(2n) \).
The maximally symmetric case is the one in figure \ref{cechs} A. In the original
theories (and with this we mean the theories realized as configurations of branes
with orientifold planes) the \( SO(2n) \) subgroup symmetry is manifestly realized
as the symmetry that rotates the flavors transforming in the fundamental representations
of the gauge groups; in general, this symmetry is reduced to \( SO(2n_{1})\times SO(2n_{2}) \).
The \( USp(2) \) symmetry instead is not manifest classically, as standard
in mirror symmetry \cite{intseib}; it makes its appearance as an IR symmetry,
and, in general, is reduced to \( U(1) \). Alternatively, if one turns on a
magnetic gauge coupling, the symmetry is present only at infinite magnetic gauge
coupling. In the mirror theories instead the \( SO(2n) \) subgroup symmetry
is not manifest classically, while the \( USp(2) \) symmetry (or, generically,
its \( U(1) \) subgroup) rotates the flavors corresponding to external lines.
These symmetries can be also read from the brane configurations as the symmetries
on the world-volume of NS- and D5-branes. All these symmetries are pictorially
manifest in the quiver diagram of the mirror theory; \( SO(2n) \) is associated
with the \( D_{n} \) Dynkin-diagram form of the graphs and \( USp(2) \) rotates
the external lines. The number and the position of the external lines also accounts
for the reduction of this symmetry group to a generic \( U(1)\times SO(2n_{1})\times SO(n_{2}) \)
subgroup. 

A natural question is: what is the mirror of the theory corresponding to a \( D_{n} \)
quiver diagram with up to \( k \) external lines arbitrarily distributed among
the nodes? The answer is not difficult to find using the results in the previous
sections: we need to consider the case with more NS-branes and make a diagrammatic
computation with the rules we discussed. The result will be presented in section
 \ref{zeta}. But, first, let us make some general remarks and try to make contact
with different approaches.

\subsection{Discussion and relation to other approaches \label{M}}

The step-by-step rules for computing mirror pairs, which we discussed in the
previous sections, can be applied to a large variety of theories. In this section,
we discuss few general issues about our results and consider examples that have
a T-dual description and allow to make contact with different approaches. 

All the previous pairs of mirror theories pass the simplest consistency check:
the dimension of the Coulomb branch of the theory is equal to the dimension
of the Higgs branch of its mirror, and vice versa. The number of masses and
FI terms is also consistent with the mirror symmetry expectations, except that
in the case of the theories depicted in figure \ref{cechs}. The \( USp(2k)^{2} \)
theory for \( n_{1}=0,1 \) indeed has a missing FI term. A similar phenomenon
was encountered in \cite{pz} and also in \cite{kapu}, in the analysis of \( D_{2} \)
and \( D_{3} \) quiver diagrams. A possible resolution of this paradox is that
the \( USp(2k)^{2} \) theory has a hidden FI parameter which is not visible
in the classical Lagrangian and appears only when the theory flows in the IR
to an interacting superconformal fixed point. 

The reader may have noticed that the theories \( USp(2k)^{2} \) and \( U(2k) \)
with antisymmetric tensors can be interpreted as the world-volume theory of
\( SO(2n) \) small instantons sitting at a \( Z_{2} \) singularity \cite{intblum1,intblum2,hzsix}.
The representation of these small instanton theories as systems of NS-, Dp-
and Dq-branes was extensively discussed in \cite{hzsix, karchbr}; a T-duality
along \( x_{6} \) transforms the NS-branes into a space-time singularity, and
the configuration of D-branes into a system of D2- and D6-branes, which have
a natural interpretation as small instanton theories. In this approach the orientifold-invariant
dispositions of two NS-branes on a segment accounts for all the possible breaking
of \( SO(2n) \) to subgroups and the type of allowed bundles (with or without
vector structure) \cite{hzsix}. Using the discussion in section  \ref{T},
it is not difficult to see that also the mirror theories have the natural interpretation,
in the spirit of \cite{hzsix}, as the world-volume theory of \( SU(2) \) small
instantons sitting at a \( D_{n} \) singularity. Once again, modulo minor differences
due to the presence of \( ON^{0} \), the disposition of the NS-branes on the
segment accounts for the allowed global symmetry bundle. We see that the gauge
group of the instanton theory and the group associated to the singularity are
exchanged by mirror symmetry. This is not a fortuitous coincidence. 

In \cite{pz} mirror symmetry was studied by realizing the gauge theories with
configurations of D2- and D6-branes, and lifting them to systems of membranes
sitting at orbifold singularities in M theory. This approach can be useful for
studying the N=4 three-dimensional superconformal fixed points using the tools
of the AdS/CFT correspondence \cite{fkpz,gomis}. The compact examples considered
in the previous sections can be reduced to configurations of D2- and D6-branes
by performing a T-duality along \( x_{6} \) and using the results of section
 \ref{T}. Consider only the case in which there is the same number of D3-branes
everywhere; more general configurations are associated with fractional branes
sitting at orbifold singularities. We obtain systems of D2-branes sitting at
a \( \Gamma _{G} \) singularity in the presence of D6-branes realizing a \( G' \)
global symmetry, where \( G \) and \( G' \) are \( SU(n) \) or \( SO(2n) \);
these theories are the world-volume theories of small \( G' \) instantons sitting
at a \( G \) singularity. Mirror symmetry correspond to the exchange of \( G \)
with \( G' \). The system can be lifted to a set of membranes in M-theory with
a \( G\times G' \) singularity; mirror symmetry is then reduced to the geometrical
symmetry that exchanges \( G \) with \( G' \). The argument is general and
simple; unfortunately, the three-dimensional theory is not completely specified
until we specify the form of the \( G' \) bundle on the \( \Gamma _{G} \)
ALE space, since in general the global symmetries \( G \) and \( G' \) are
broken to subgroups by the choice of a bundle. The case analyzed in the previous
sections corresponds to the singularity \( Z_{2}\times D_{n} \). The mutual
disposition of NS- and D5-branes in both the original and the mirror theory
determines a particular bundle, with a given global symmetry; in the case of
the D-type quiver diagram, the disposition of the external lines gives a pictorial
description of this bundle. The Type IIB description is useful for describing
the most general bundle, and, more than that, provides a step-by-step method
for computing the bundle for the mirror theory once the original one is given.
Moreover, the type IIB description can be easily used also for configurations
containing fractional branes. 

Finally, we notice that interesting effects are associated with the singular
cases of \( D_{2} \) and \( D_{3} \) singularities. In this paper we do not
discuss these cases; however, they are easily realized using the methods discussed
in this Section. The case of a \( D_{2} \) singularity is particularly interesting
since it provides examples of theories that are dual (in the sense that they
flow to the same IR fixed point) at specific points in moduli space \cite{sethi, kapuber}.
The reason is that a \( D_{2} \) singularity splits into two \( A_{1} \) singularities.
The M-theory lift of two D6-branes near an orientifold is the space \( (R^{3}\times S^{1})/Z_{2} \).
The two fixed points of \( Z_{2} \) correspond, in general, to different superconformal
fixed points and the three-dimensional gauge theory may flow to one or the other
according to the vacuum expectation value of the scalar parameterizing the position
in \( S^{1} \). It is not immediately obvious how these effects can be seen
in the Type IIB description we are using in this paper.

\subsection{The case with \protect\( Z_{m}\times D_{n}\protect \) global symmetry\label{zeta}}

\begin{figure}
{\par\centering \resizebox*{16cm}{!}{\includegraphics{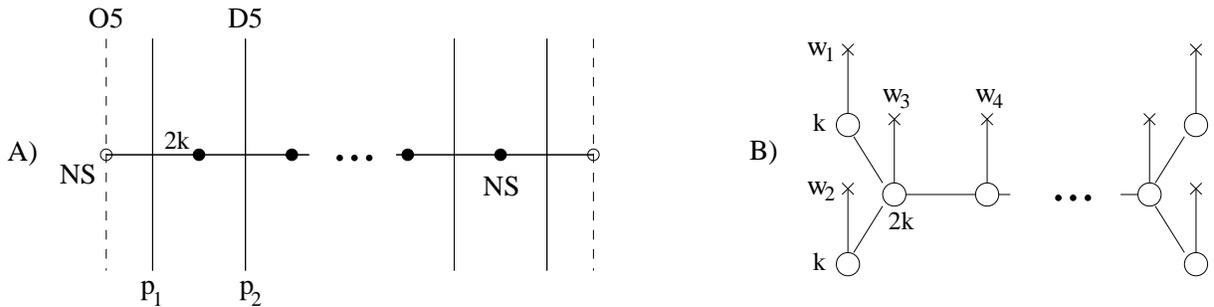}} \par}

\caption{In figure A there are a total of \protect\( \sum _{i}p_{i}=n\protect \) D5-branes
and a total of \protect\( m\protect \) NS branes (carrying \protect\( \frac{m}{2}\protect \)
magnetic NS charge). The NS-branes are put on the segment in an orientifold-invariant
way. Stuck NS-branes on the orientifolds can be present or not; this has been
denoted using unfilled circles. The theory is \protect\( USp(k)\times U(k)^{\frac{m}{2}-1}\times USp(k)\protect \)
with bi-fundamentals and \protect\( n\protect \) fundamentals distributed among
the gauge factors, or analogous theories in which one or both the \protect\( USp(k)\protect \)
factors are replaced by \protect\( U(k)\protect \) with an antisymmetric tensor,
if there are one or two stuck NS-branes, respectively. The mirror theory is
depicted in figure B; the \protect\( D_{n}\protect \) quiver diagram has a
maximum number \protect\( m\protect \) of external lines. \label{lastIhope}}
\end{figure}

In this section we consider examples with global symmetry of type \( Z_{m}\times D_{n} \).
In the original configurations of branes constructed with orientifolds, these
models are obtained by adding more NS-branes to the examples considered in section
 \ref{examples}; in terms of the mirror theories, they are obtained by adding
external lines to the \( D_{n} \) quiver diagram. 

Consider only compact configurations. Non compact ones, which end with a NS-brane,
can be extended to compact one by adding a second orientifold plane. As it should
be clear from the examples in section  \ref{examples}, the mirror transform
of a non-compact model can be obtained by the mirror transform of the corresponding
compact one, by substituting the right part of the quiver diagram with a standard
pattern of nodes associated to groups with decreasing rank, as depicted in figure
\ref{spdynkin}. 

We consider only, for simplicity, configurations with the same number of D3-branes
everywhere. These are the kind of configurations that have a known T-dual description.
The reader has all the elements for analyzing more complicated examples. 

Consider figure \ref{lastIhope} A. \( m \) NS-branes can be put on the segment
in an orientifold-invariant fashion in several ways \cite{hzsix, karchbr}.
If \( m \) is odd, there must be a stuck NS-brane; each of the \( [m/2] \)
physical branes living not at the boundary of the segment has an image under
the orientifold projection. If \( m \) is even, we can put \( \frac{m}{2} \)
physical branes on the segment far from the boundary (case with vector structure)
or we can put two stuck and \( \frac{m}{2}-1 \) physical NS-branes (case without
vector structure). The theory is \( USp(k)\times U(k)^{\frac{m}{2}-1}\times USp(k) \)
with bi-fundamentals and \( n \) fundamentals distributed among the gauge factors,
or analogous theories in which one or both the \( USp(k) \) factors are replaced
by \( U(k) \) with an antisymmetric tensor; the replacement occurs if there
are one or two stuck NS-branes, respectively. These theories represent small
\( SO(2n) \) instantons sitting at a \( Z_{m} \) singularity \cite{intblum1,intblum2,hzsix}. 

The mirror is a \( D_{n} \) quiver theory with an arbitrary distribution of
a maximum of \( m \) fundamentals among the nodes. The exact mirror can be
easily derived using the rules discussed in section  \ref{examples}.

These pairs of mirror theories are associated to M-theory singularities of the
form \( Z_{m}\times D_{n} \). These theories have been also considered in \cite{pz}
in the simplified case in which \( k=1 \) and the maximal global symmetry is
not broken \footnote{
The proposal in \cite{pz}, when compared to the one in this section, has an
interchange between theories with vector structure and theories without. This
happened since, in those days, there was no explicit method for computing the
mirror theories in the case of \( D_{n} \) global symmetries, and the result
was guessed on the basis of the M-theory intuition, the symmetry of the problem
and the counting of parameters. In this paper we presented a method for explicitly
computing the mirror theory.
}. The discussion in this section provides a general method for studying generic
bundles and for determining the particular theory that corresponds to a generic
diagram with a given distribution of external lines among the nodes. 

The discrepancy between the number of parameters of the theories and of their
mirrors mentioned in section  \ref{M} is even increased for \( k>2 \). Some
FI terms are missing; this was also noticed in \cite{pz}.

\subsection{The \protect\( D_{n}\times D_{k}\protect \) global symmetry and self-dual
models}

\label{section:DD}

We finish our discussion about mirror symmetry by considering the class of theories
realized using both \( ON^{0} \) and orientifolds planes. Consider, for example,
figure \ref{example}. The three dimensional gauge theories that can be realized
using these configurations are encoded in the quiver diagrams shown in the figure.
We may ask: what is the strong coupling limit of these configurations? Luckily
enough, the answer is very simple; since \( ON^{0} \) and \( O5^{0} \) are
exchanged by S-duality, the mirror configuration has a form similar to the original
one. It is encoded in a quiver diagram of the same form, with different numbers
\( w_{i} \) and different dispositions of external lines. Consider only, for
simplicity, compact models and configurations with the same number of D3-branes
everywhere.
\begin{figure}
{\par\centering \resizebox*{10cm}{!}{\includegraphics{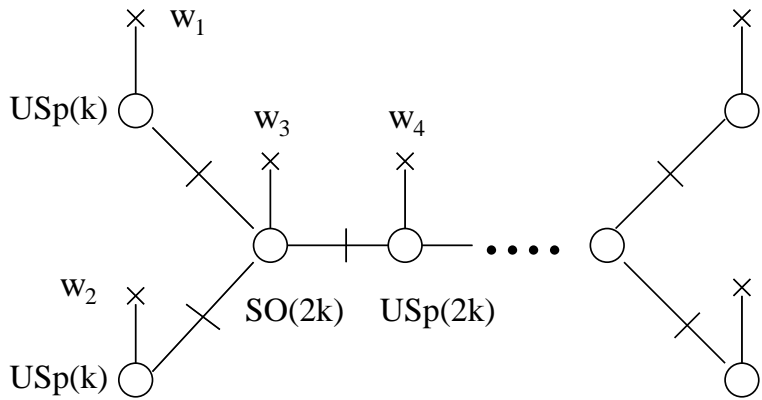}} \par}

\caption{A \protect\( D_{n}\protect \) quiver diagram with \protect\( k=\sum w_{i}\protect \)
external lines. It corresponds to a global symmetry associated to \protect\( D_{n}\times D_{k}\protect \).
The mirror theory has the same form, with \protect\( k\protect \) and \protect\( n\protect \)
interchanged and with a different partition of \protect\( n=\sum \tilde{w}_{i}\protect \)
external lines.\label{DD}}
\end{figure}

The maximally allowed global symmetry is \( SO(2n)\times SO(2k) \), where \( k=\sum w_{i} \).
The disposition of external lines breaks this global symmetry to a subgroup
\( SO(w_{1})\times SO(w_{2})\times USp(w_{3})\times USp(w_{4})\times \cdots  \).
This is the world-volume theory of an \( SO(2k) \) small instanton sitting
at a \( D_{n} \) singularity. The corresponding M-theory singularity is \( D_{n}\times D_{k} \).
Using the explicit S-duality, with the rules explained in the previous sections,
or using the arguments in \cite{pz}, we conclude that the mirror theory has
the form in figure \ref{DD}, but with \( k \) and \( n \) interchanged. The
precise number and disposition of external lines can be determined by an explicit
(and easy) computation. 

It is natural to consider self-dual examples. If \( k=n \), choosing suitable
values for \( w_{i} \), we easily obtain self-dual theories. In other words,
we can consider these configurations as obtained by the orbifold generated by
\( \{(-1)^{F_{L}}R_{6789},\Omega R_{3456}\} \); this is the meaning of having
both \( ON^{0} \) and \( O5^{0} \), as discussed in section \ref{section: oointer}.
Under an S-duality, \( \Omega  \) and \( (-1)^{F_{L}} \) are interchanged.
The orbifold is therefore unaffected by an S-duality combined with the interchange
of \( (345) \) with \( (789) \) (this are the rules of the game \cite{hw}!);
configurations in which the total D5-charge and the total NS-charge, as well
as their disposition in space-time, are the same (this certainly implies \( k=n \))
are therefore expected to be self-dual.

\subsection{Open Problems}

There are open questions which are not solved by the current methods we use.
There are two basic unsolved questions. 

The mirror of SO gauge theories and their generalizations. We refer to configurations
which involve \( O3^{+} \) or \( O5^{+} \) planes. It is not clear at the
moment how to understand the mirror of these planes and this is the source for
the confusion. 

There are two basic constructions of \( USp \) gauge theories using D3 branes,
D5 branes and NS branes. One construction uses an \( O3^{+} \) plane and the
other one an \( O5^{-} \) plane. In both cases the resulting gauge group is
\( USp \). Correspondingly, the Type IIB S-dual of both configurations is expected
to give the known dual. Currently only one configuration gives a satisfactory
answer. The one with an \( O5^{-} \) plane. The configuration with \( O3^{+} \)
is not understood well enough to produce the right mirror.

\section{Four Dimensional Theories \label{fourd}}

\subsection{Brane Box Models}

\begin{figure}
{\par\centering \resizebox*{15cm}{!}{\includegraphics{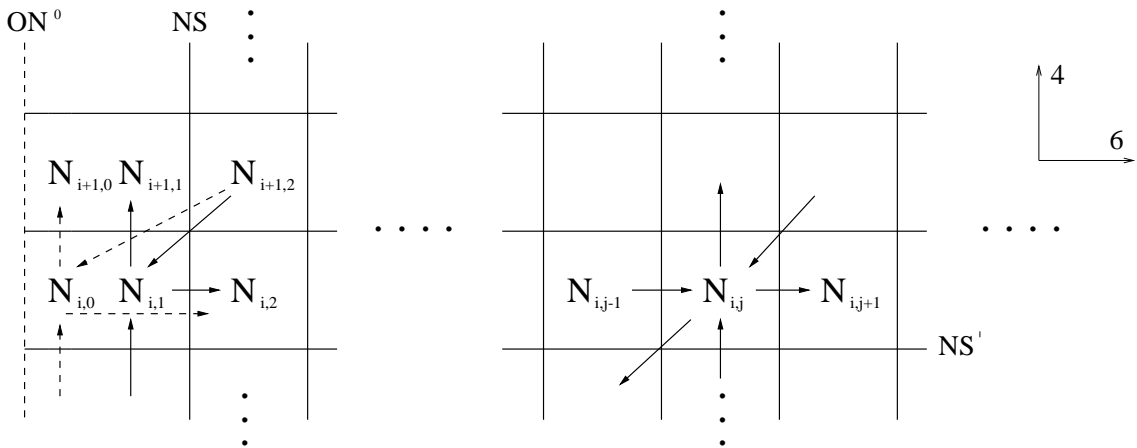}} \par}

\caption{A generic Brane Box Model with \protect\( ON^{0}\protect \) planes. The matter
fields in the fundamental or anti-fundamental representation, which are represented
as oriented arrows, are indicated for two particular boxes. The boxes near \protect\( ON^{0}\protect \)
are split into two \protect\( U(N)\protect \) gauge factors. The matter fields
for \protect\( U(N_{i,1})\protect \) are indicated as standard arrows, while
those for \protect\( U(N_{i,0})\protect \) are indicated as dashed arrows.
If \protect\( N_{i,j}=2N\protect \) for \protect\( j\neq 0,1\protect \) and
\protect\( N_{i,0}=N_{i,1}=N\protect \), there is no bending and the one-loop
beta function is zero. Under the same condition, if the model is compact in
both the horizontal and vertical direction, the theory is finite and related
under T-duality to D3-branes near an orbifold of \protect\( C^{3}\protect \)
generated by \protect\( Z_{k}\protect \) and \protect\( D_{n}\protect \),
where \protect\( k\protect \) is the number of NS\protect\( '\protect \)-branes
and \protect\( n-2\protect \) is the number of NS-branes.\label{box}}
\end{figure}

It is almost straightforward to introduce an \( ON^{0} \) plane in the Brane
Box Models studied in \cite{hzfour}. Consider figure \ref{box}. There are
D5-branes extending in \( (012346) \), NS-branes and \( ON^{0} \) planes extending
in \( (012345) \) and NS\( ' \)-branes extending in \( (012367) \). The model
has \( N=1 \) supersymmetry and is generically chiral. The generic box \( N_{i,j} \)
gives rise to the gauge group \( U(N_{i,j}) \). The open strings connecting
neighboring boxes give rise to chiral matter fields in the fundamental (or anti-fundamental)
representation of the neighboring gauge groups. These matter fields and their
chirality are depicted using oriented arrows. Only arrows directed and oriented
East, North and South-West exist \cite{hzfour}. All the existing matter fields
are indicated for the box \( N_{i,j} \). Every time three arrows close a triangle
there is a cubic superpotential \cite{hzfour}. The only novelty regards the
boxes near \( ON^{0} \), for example \( N_{i,0} \) and \( N_{i,1} \). As
we can expect from section  \ref{onplane}, there is a splitting into two gauge
factors \( U(N_{i,0}) \) and \( U(N_{i,1}) \). The action of \( (-1)^{F_{L}} \)
on the Chan-Paton factors is a diagonal matrix with entries \( +1 \) for the
indices associated to \( N_{i,0} \) and \( -1 \) for those associated to \( N_{i,1} \).
As a consequence, the open strings connecting, say, \( N_{i,0} \) with \( N_{i\pm 1,1} \)
are projected out. The resulting spectrum is indicated in figure \ref{box}
for the box \( N_{i,1} \) and \( N_{i,0} \); the standard arrows represent
the matter fields for the gauge group \( U(N_{i,1}) \), while the dashed arrows
represent the matter fields for \( U(N_{i,0}) \). 

The numbers \( N_{i,j} \) must be chosen in order to have an anomaly-free model.
Anomalies on the world-volume of the branes should be related to the violation
of the equations of motion for some space-time field. Due to the complexity
of the model and the non-trivial bending of the NS and NS\( ' \)-branes, the
precise relation between anomalies and charge conservation of the string background
is not known. Attempts to find such a relation appeared in the literature \cite{list1,list2,list3,list4}
without a conclusive result. In principle, it may happen that, even if the number
\( N_{ij} \) are such that they cancel anomalies, the branes background is
inconsistent. However, the consistency of a large class of Brane Box Models
can be explicitly checked by T dualizing them to other consistent systems of
branes sitting at orbifold singularities \cite{hu,list3}. The class of consistent
models can be enlarged by considering models which are separately well defined
and sewing them \cite{list1,list2,list3,list4}. 

There is an obvious example that is well defined. If we take the same number
of D5-branes in each box there is no bending for the NS-branes and the space-time
equations of motion are satisfied. This corresponds to \( N_{i,j}=2N \) for
\( j\neq 0,1 \) and \( N_{i,0}=N_{i,1}=N \). The condition of no bending is
equivalent to vanishing of the total charge of the D5-branes ending on a given
NS and NS\( ' \)-branes. The analogous condition for the \( ON^{0} \) plane
forces us to take \( N_{i,0}=N_{i,1} \). Since the bending is associated to
the running of the coupling constant \cite{witten2}, this model has zero one-loop
beta function \cite{hstrassu}. Every gauge group \( U(n) \) , including those
realized near \( ON^{0} \), has indeed \( 3n \) fields in the fundamental
and \( 3n \) in the anti-fundamental representation. The quantum field theory
is conjectured to be finite using the same argument of non-bending as in \cite{hstrassu}.

\subsection{Relation to Branes at Orbifold Singularities and Finite Models}

We can easily construct cylindrical (6 direction compact) and elliptical (6
and 4 directions compact) models. These kind of models can be related by T-duality
to systems of branes at orbifold singularities \cite{hzfour, hstrassu, hu}.
Consider \( n-2 \) NS-branes and \( k \) NS\( ' \)-branes. According to the
discussion in section  \ref{T}, we expect that the NS\( ' \)-branes are replaced
by a T-duality along the \( 4 \) direction with an A-type singularity and the
NS-branes and the \( ON^{0} \) planes are replaced by a T-duality along the
6 direction with a D-type singularity. The cylindrical models are the T dual
of systems of D4 and NS-branes at a \( D_{n} \) orbifold singularity extended
in the directions \( 6,7,8,9 \) \cite{hzfour, poppitz}. The elliptical models
are the T dual of systems of D3-branes sitting at a \( C^{3} \) orbifold; the
actual orbifold group is the smallest discrete group acting on \( C^{3} \)
that contains a \( Z_{k} \) subgroup acting on the coordinates \( 4,5,8,9 \)
and a \( D_{n} \) subgroup acting on the coordinates \( 6,7,8,9 \). To preserve
\( N=1 \) supersymmetry, this discrete group must be a subgroup of \( SU(3) \)
\cite{hevariste}. It would be quite interesting to further analyze the relation
between the Brane Box Models and the construction in \cite{hevariste}. 

When \( N_{i,j}=2N \) for \( j\neq 0,1,n,n-1 \) and \( N_{i,0}=N_{i,1}=N_{i,n-1}=N_{i,n}=N \),
the D3-branes transforms in the regular representation of the orbifold group
\cite{hm}. However, in general, the numbers \( N_{i,j} \) do not need to be
all equal. We may project the D3-branes with a different representation. This
would result in having fractional D3-branes. The anomaly cancellation becomes
now equivalent to the tadpole cancellation \cite{list3}. 

We may obtain more complicated orbifold singularities by compactifying in the
\( 6 \) direction allowing for a shift along the \( 4 \) direction \cite{hu}.
The shift must be compatible with the \( Z_{2} \) projection induced by \( ON^{0} \).
It appears that the Brane Box Models give a simple construction of the gauge
theory and the superpotential of some of the theory associated with discrete
groups of \( SU(3) \). It would be interesting to analyze the dictionary for
translating these kind of models into an orbifold projection, along the lines
of \cite{hu, hevariste}. 

It is believed that the properties of finiteness improve if the models are compact
\cite{hstrassu,hu}. If we take \( N_{i,j}=2N \) for \( j\neq 0,1,n,n-1 \)
and \( N_{i,0}=N_{i,1}=N_{i,n-1}=N_{i,n}=N \) and compactify along the \( 6 \)
direction, allowing for some shift, we obtain a finite and conformal \( N=1 \)
model. As we said, this is mapped by T-duality to a set of D3-branes sitting
at an orbifold singularity. The finiteness of this model then follows from the
AdS/CFT correspondence using the arguments in \cite{eva}.

\section{Six-Dimensional Theories\label{section: sixd}}

In this section  we discuss examples of six-dimensional gauge theories. 

There are several consistent string backgrounds that give rise to anomaly-free
six-dimensional theories \cite{augusto1,augusto2,augusto3,gp,gj,intblum1,intblum2}.
In \cite{hzsix} we described a realization with D6-, D8- and NS-branes. The
introduction of an \( ON^{0} \) plane in the original construction in \cite{karch,hzsix,karchbr}
allows to realize more complicated examples. This section  completes the results
in \cite{hzsix}. It also shows how to construct (in the spirit of \cite{hw}
) the gauge theories associated to small \( SO(32) \) instantons sitting at
\( D_{n} \) singularities. 

In section    \ref{onplane}, we discuss in detail what is the world-volume
theory of Dp-branes ending on NS-branes and \( ON^{0} \) planes. In the case
of D6-branes, we must include the fields living on NS-branes or \( ON^{0} \)
planes in the six-dimensional theory and we must also pay attention to the anomaly
cancellation conditions. As a general rule, an anomaly in field theory translates
in the brane set-up to the non-conservation of some charge for bulk fields.
In this particular case, the relevant charges are associated to the fields living
on the NS-branes; for a remarkable return, the same fields provide the tensor
multiplets that are necessary to completely cancel the anomalies in six-dimensions. 

The general discussion about anomaly cancellation in the brane set-up can be
found in \cite{hzfour2,hzsix}. The general rule will be clear after considering
a specific example. The six-dimensional theory corresponding to figure \ref{rule}B
is still \( U(n_{1})\times U(n_{2}) \) without bi-fundamentals, but now with
two tensor multiplets and two hyper-multiplets coming from the NS-brane and
the \( ON^{0} \) plane. Naively, one would think that only one linear combination
of these multiplets is relevant for the field theory, arguing that the sum of
the two multiplets decouples while the difference appears as a gauge coupling
and FI parameter, respectively, for both gauge groups. However, this is not
true. The two tensor multiplets and two hypermultiplets couple both to the gauge
theory. The sum of the multiplets couples to, say, the gauge group on the D6
branes which have positive charged with respect to the \( ON^{0} \) multiplet
while the difference of the multiplets couples to the gauge group associated
with the negatively charged D6 branes. The theory in figure \ref{rule} is generally
anomalous. A \( U(N) \) theory with \( N_{f} \) flavors indeed is anomalous
unless \( N_{f}=2N \) and the theory is coupled to a tensor multiplet, the
scalar of which plays the role of a gauge coupling. It is easy to see what is
wrong with charge conservation for bulk fields. Since the world-volume of the
D6-brane is bigger than that of a NS-brane, the RR-charge of the D6-brane can
not be absorbed by the NS-brane as it happens for all the Dp-branes ending on
a NS-brane for \( p\leq 5 \). Therefore the D6 charge must be canceled locally
at the position of the NS-brane. In this case, there are \( n_{1}+n_{2} \)
D6-branes on the left of the NS-brane and zero on the right and the charge is
not conserved\footnote{
Notice that, while a Dp-brane may have positive or negative charge under the
fields living on \( ON^{0} \) \cite{senS}, the charge under the NS-brane fields
is always positive for branes ending, say, on the left. We are not considering
anti Dp-branes in our picture; they would break supersymmetry. Actually, one
can interpret Sen's assignment of positive and negative charges as being ending
`to the left' or `to the right' of the \( ON^{0} \) plane, respectively as
in figure \ref{LAST}. 
}.
\begin{figure}
{\par\centering \resizebox*{17cm}{!}{\includegraphics{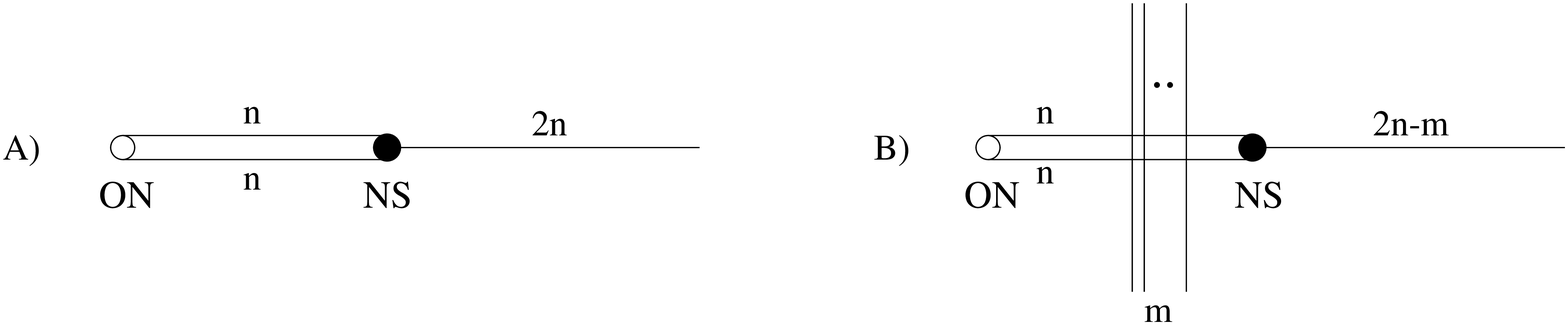}} \par}

\caption{The standard figure for a six-dimensional theory. D6-branes are depicted as
horizontal lines ending on points corresponding to NS-branes (filled circles)
and \protect\( ON^{0}\protect \) planes (empty circles). D8-branes are depicted
as vertical lines. \label{sixrule}}
\end{figure}

There are two ways of removing the obstacle and constructing an anomaly-free
theory. They are depicted in figure \ref{sixrule}. Notations are changed with
respect to figure \ref{rule}. Since the D6-worldvolume is bigger than the NS
one, it is better to represent D6-branes as lines ending on points which represent
NS-branes. In figure A, we added \( n_{1}+n_{2} \) semi-infinite D6-branes
to the right of the NS-brane in order to preserve the charge. The two kinds
of D6-branes ending on \( ON^{0} \) have opposite charge with respect to the
fields living on it and we must cancel the \( ON^{0} \) tadpole by taking \( n_{1}=n_{2} \).
The resulting theory is \( U(n)\times U(n) \) with \( 2n \) flavors for each
group, two tensor multiplets and two hypermultiplets. The tensor multiplets
are necessary to cancel the non-Abelian anomalies \cite{sagnotti} while the
hypermultiplets, acting as dynamical FI terms, cancel the Abelian anomalies,
making the \( U(1) \) factors massive \cite{hm,leigh}. In figure B, we also
added D8-branes, depicted as vertical lines. They induce a cosmological constant,
which is constant in space-time and jumps by one unit when one crosses a D8-brane
\cite{wpol}. A non-zero cosmological constant \( m \) induces an effective
RR seven form charge at the position of a NS-brane. This changes the charge
conservation condition in the following way: the D6 charge on the left of a
NS-brane minus the D6 charge on the right must equal the value of the cosmological
constant at the position of the NS-brane\footnote{
We are using conventions in which a physical D6-brane counts \( +1 \) and a
physical D8-brane induces a cosmological constant of magnitude \( +1 \). 
}: \( n_{l}-n_{r}=m \) \cite{hzfour2,hzsix}. The cosmological constant is zero
near \( ON^{0} \) implying again \( n_{1}=n_{2} \). It is instead equal to
\( m \) near the NS-brane implying that we must add \( 2n-m \) semi-infinite
D6-branes to its right. The resulting theory is again \( U(n)\times U(n) \)
with \( 2n \) flavors for both groups, two tensor multiplets and two hypermultiplets. 

Equipped with the rules discussed in section  \ref{onplane}, we can construct
more general examples. Adding NS-branes to the initial \( ON^{0} \) plane,
without including orientifolds, we obtain theories that are products of \( U(N) \)
gauge groups. This kind of examples are exhaustively discussed in \cite{hzsix}. 

Introducing an \( O8 \) plane, we can construct many more examples, containing,
in particular, the configurations T dual to \( SO(w) \) small instantons on
\( D_{n} \) ALE spaces. 

Compact model with two \( ON^{0} \) planes are associated to small instanton
theories, as discussed in section  \ref{T}. In the six-dimensional case, we
must pay attention to the anomaly cancellation, which usually constrains the
world-volume theory. Theories of small \( U(N) \) or \( SO(2N) \) instantons
living at a \( D_{n} \) singularity are associated with a \( D_{n} \) extended
Dynkin diagrams as in figure \ref{braneB} with the addition of external lines
as in figure \ref{example1}, for \( U(N) \), or figure \ref{example}, for
\( SO(2N) \). The \( n \) tensors and hypermultiplets required to cancel both
non-Abelian and Abelian anomalies are automatically provided by the \( n-2 \)
NS-branes and the two \( ON^{0} \) planes. 

The introduction of an \( O8^{-} \) plane affects our picture much more than
in the original examples in \cite{hzsix}. The general strategy in \cite{hzsix}
was to replace every space-time singularity with a dual background where the
singularity was replaced by branes. This in general allows to have a better
control on the parameters and moduli of the theory. In the case of \( D_{n} \)
singularities, we will have a mixed picture, with both NS-branes and the perturbative
orbifold projection associated to \( ON^{0} \) plane.
\begin{figure}
{\par\centering \resizebox*{8cm}{!}{\includegraphics{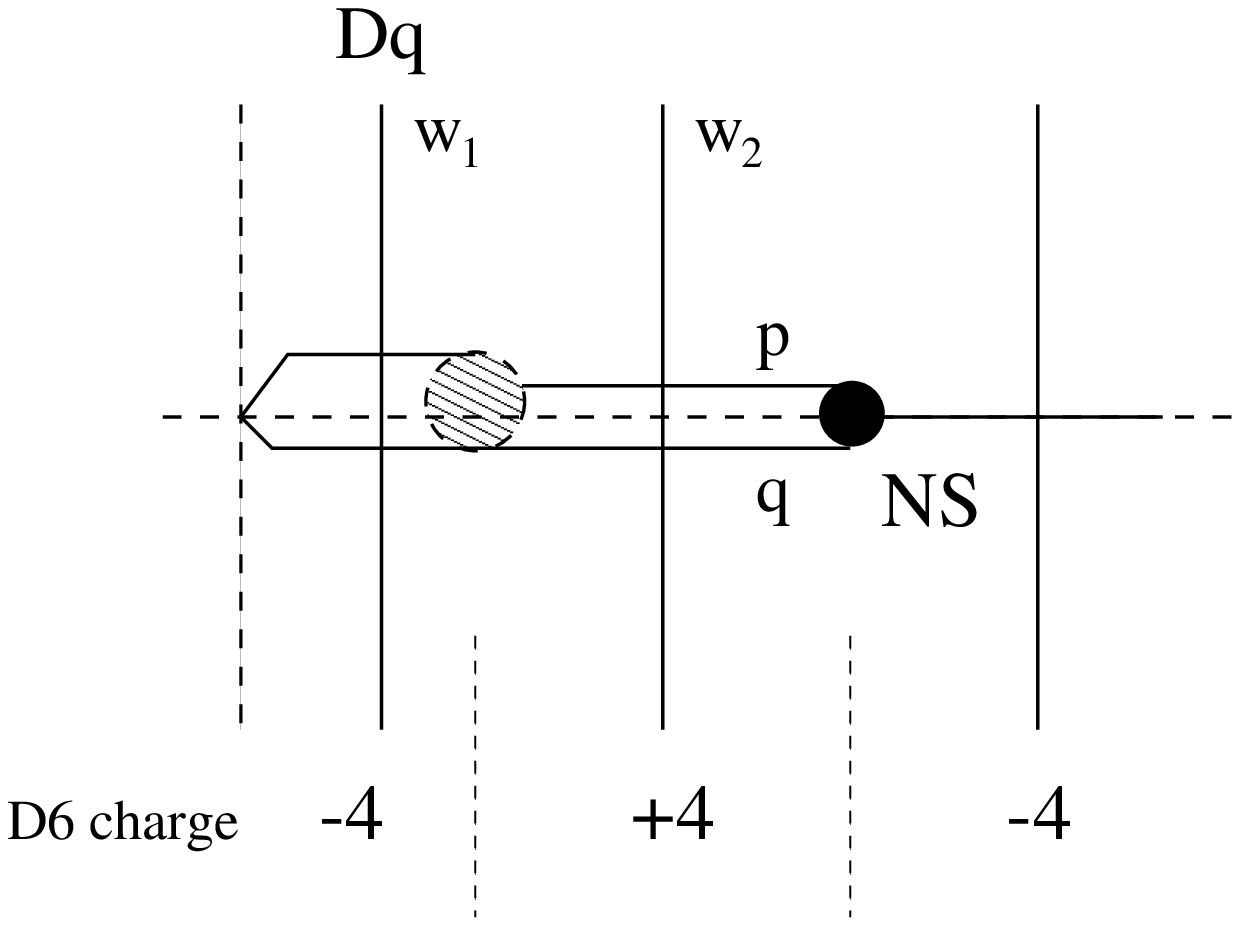}} \par}

\caption{The anomaly cancellation rule near \protect\( ON^{0}\protect \). \protect\( ON^{0}\protect \)
is represented as the pair containing a fixed plane and a virtual NS-brane.\label{FINAL}}
\end{figure}

The generic quiver theory in the presence of \( O8^{-} \) plane was already
discussed, for arbitrary dimension, in section \ref{section: oointer} (figure
\ref{example}). The form of the world-volume theory does not depend on the
dimension. What is peculiar to six-dimensions is that the anomaly cancellation
uniquely determines the world-volume theory. Moreover we need to include fields
from the NS-branes and from the twisted sectors in the theory; they are crucial
for canceling the anomalies. Recall from section  \ref{section: oointer}, that
there are two consistent configurations of D-branes living at the intersection
 of an \( O8^{-} \) and an \( ON^{0} \) plane. The first one has an \( O6^{-} \)
plane, no splitting of branes, \( U \)-type gauge groups and an hypermultiplet
from the twisted states. The second one has an \( O6^{+} \) plane, a splitting
with \( USp \)-type gauge groups and a tensor multiplet from the twisted states.
In six-dimensions, it is crucial to keep track of the twisted sectors, since
they contribute to the theory and are important in canceling the anomalies. 

The actual gauge groups (the numbers \( p,q,t,s,... \) in figure \ref{example})
can be determined by the RR-charge conservation condition, which is equivalent
to the anomaly cancellation \cite{hzsix}. We recall that we are using notations
where \( O8^{-} \) contributes \( -8 \) to the cosmological constant and \( O6^{\pm } \)
has RR-charge \( \pm 4 \). The only new ingredient, not present in \cite{hzsix},
is the condition that must hold near \( ON^{0} \). The total charge for the
\( ON^{0} \) twisted tensor field must be canceled, since there is no room
for the D6-brane RR-charge to escape. This fixes the splitting (the numbers
\( p \) and \( q \)) near \( ON^{0} \). The rule is that \( p-q=w_{1} \).
It is easily checked that it is equivalent to the anomaly cancellation condition;
we give an explicit example below. This result would surely follow from a careful
analysis in terms of boundary states, along the lines of \cite{senS}. We prefer
to give a heuristic argument in the spirit of the S-dual picture of figure \ref{LAST}.
It is convenient to represent the \( ON^{0} \) plane as splitted into a fixed
plane and a \textit{virtual} NS-brane which supports the twisted fields (see
figure \ref{FINAL}). The \( w_{1} \) Dq-branes now live on the left of the
virtual NS-brane. We assume that the virtual NS-brane behaves at all effects
as a real NS-brane. This means that the orientifold plane \( O6 \) change sign
in crossing it. More important, the anomaly cancellation condition for the \( ON^{0} \)
plane is mapped to the standard condition for a NS-brane. The cosmological constant
at the virtual NS-brane is \( -8+w_{1} \), where \( -8 \) is the contribution
of \( O8^{-} \). The condition is \( (2p-4)-(p+q+4)=-8+w_{1} \), which is
the same as \( p-q=w_{1} \); the \( \pm 4 \) are the contribution of the \( O6 \)
plane on the right and on the left of the virtual NS-brane, respectively.
\begin{figure}
{\par\centering \resizebox*{12cm}{!}{\includegraphics{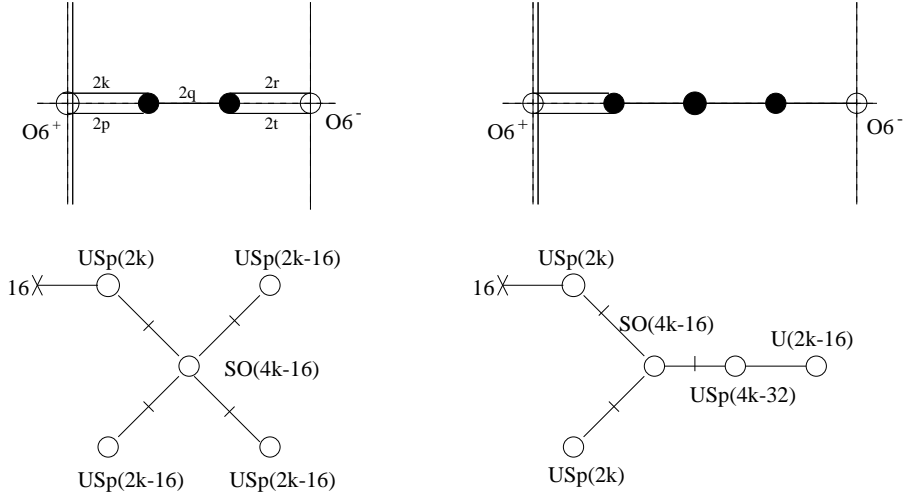}} \par}

\caption{The \protect\( SO(32)\protect \) small instanton theory for \protect\( D_{4}\protect \)
and \protect\( D_{5}\protect \) singularities. For simplicity, we considered
an \protect\( SO(32)\protect \) unbroken group, obtained by putting all the
16 D8-branes at the same point, near \protect\( ON^{0}\protect \). The case
for general breaking of the global symmetry group follows by partitioning the
D8-branes in between the NS-branes. The charge conservation rules described
in the text uniquely fix the world-volume theory. The generalization to an arbitrary
\protect\( D_{k}\protect \) is obvious.\label{small}}
\end{figure}

It is quite easy to construct compact models by putting another \( O8^{-} \)
plane in the six direction and taking \( \sum w_{i}=16 \). This will describe
the T dual version of \( SO(32) \) small instantons sitting at \( D_{n} \)
singularities \cite{intblum1,intblum2,am}. The theory will change according
to where we add the \( O8 \) plane. If \( n \) is even\footnote{
We know from section  1 that for the description of a \( D_{n} \) singularity
we need \( n-2 \) NS-branes and two \( ON^{0} \) planes. 
}, \( O8 \) will intersect an \( O6^{+} \); therefore we will need the prescription
in figure \ref{projection} B. If \( n \) is odd, \( O8 \) will intersect
an \( O6^{-} \) plane and we will need figure \ref{projection} A. The resulting
theory is the same as the one discussed in \cite{intblum1,intblum2,am}. In
figure \ref{small} we depicted the theory associated with small instantons
for \( D_{4} \) and \( D_{5} \). The general theory for arbitrary \( n \)
should be obvious from these examples; it is associated to an extended (affine)
Dynkin diagram of D-type for \( n \) even, and to a standard (without the extra
node) Dynkin diagram of D-type for \( n \) odd\footnote{
To be precise, if we consider the geometrical construction of these theories
in \cite{intblum1,intblum2}, we realize that, for \( n \) odd, it is not the
extended node of the diagram that is missing, but the last two nodes (the \( n-1 \)th
and \( n \)th) are identified. 
}. Charge conservation fixes the gauge groups in such a way to cancel all the
anomalies. As an example of the application of the previous rules, we consider
the \( D_{4} \) example in figure \ref{small}. 

We put \( w_{1}=16 \) and all the other \( w_{i}=0 \). There is a total of
four conditions. 

\begin{enumerate}
\item Near the first \( ON^{0} \); the previously discussed condition is \( 2k-2p=w_{1}=16 \). 
\item Near the first NS-brane. \( O6 \) contributes a \( +4 \) RR-charge on the
left and \( -4 \) on the right. The cosmological constant is \( -8+w_{1}=8 \),
where \( -8 \) is the contribution of \( O8^{-} \). The condition is \( (2k+2p+4)-(2q-4)=8 \). 
\item Near the second NS-brane: \( (2q-4)-(2r+2t+4)=8 \). 
\item Near the last \( ON^{0} \), where there are no D8-branes: \( 2r-2t=0 \). 
\end{enumerate}
This set of four equations determines the world-volume content indicated in
figure \ref{small}. The theory is anomaly free\footnote{
Recall that a \( USp(N_{c}) \) theory is anomaly free if \( N_{f}=N_{c}+8 \),
while an \( SO(N_{c}) \) theory is anomaly free if \( N_{f}=N_{c}-8 \). In
both cases, a tensor multiplet is needed to completely cancel the anomaly. 
}. For \( n \) even, the theory contains \( n \) tensor multiplets (\( n-2 \)
NS-branes and two \( ON^{0} \) planes intersecting \( O6^{+} \) planes); for
\( n \) odd, \( n-1 \) tensor multiplets and one FI hypermultiplet (\( ON^{0} \)
intersecting \( O6^{-} \)). This is exactly what is required to cancel the
Abelian and non-Abelian anomalies.

\subsection{Generalizations}

\begin{figure}
{\par\centering \resizebox*{8cm}{!}{\includegraphics{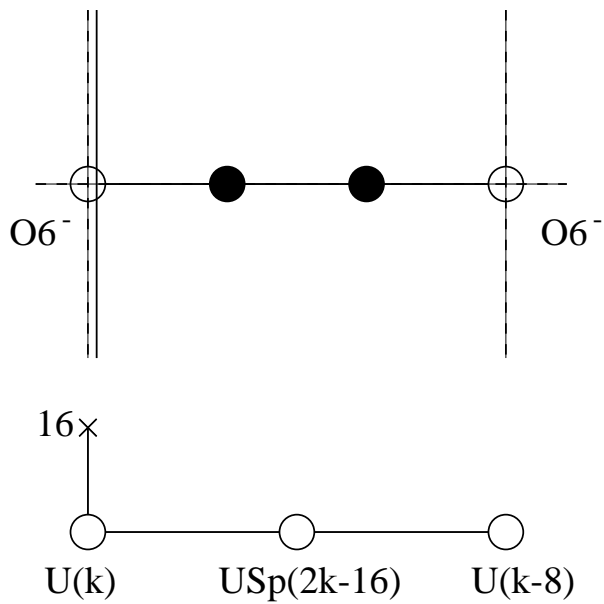}} \par}

\caption{A small instanton without vector structure sitting at a \protect\( D_{4}\protect \)
singularity. \label{without}}
\end{figure}

We can easily construct generalizations of the small instanton theories and
build new models. A first example is easily obtained by constructing theories
in which the \( ON^{0} \) plane always intersects an \( O6^{-} \) plane. There
is no splitting of branes near \( ON^{0} \) and the quiver diagram is a simple
line of nodes. This configuration is allowed only if \( n \) is even. The world-volume
theories are of the form,

\begin{equation}
\label{ex}
U(k_{1})\times USp(k_{2})\times SO(k_{3})\times \cdots \times USp(k_{n-2})\times U(k_{n-1}),
\end{equation}
 with bi-fundamentals or half bi-fundamentals for neighboring factors and fundamentals
for the various gauge groups. This model was already considered in \cite{hzsix}.
The numbers \( k_{i} \) are easily determined using the rules found in the
previous section. After orientifold projection, the two \( ON^{0} \) planes
provide the two FI hypermultiplet needed for canceling the Abelian anomalies
and the NS-branes provide the \( n-2 \) tensor multiplets needed for the non-Abelian
anomalies. We can speculate that these theories correspond to a geometrical
\( SO \) bundle at a \( D_{n} \) singularity without vector structure, which
indeed exists only for \( n \) even. We did not find the explicit construction
of theories without vector structure for D-singularities in the literature,
but it would not be too difficult to construct them using the methods in \cite{intblum1,intblum2}.
In figure \ref{without} we depicted the \( D_{4} \) example. The generalization
to \( D_{n} \) is straightforward. 

\begin{figure}
{\par\centering \resizebox*{14cm}{!}{\includegraphics{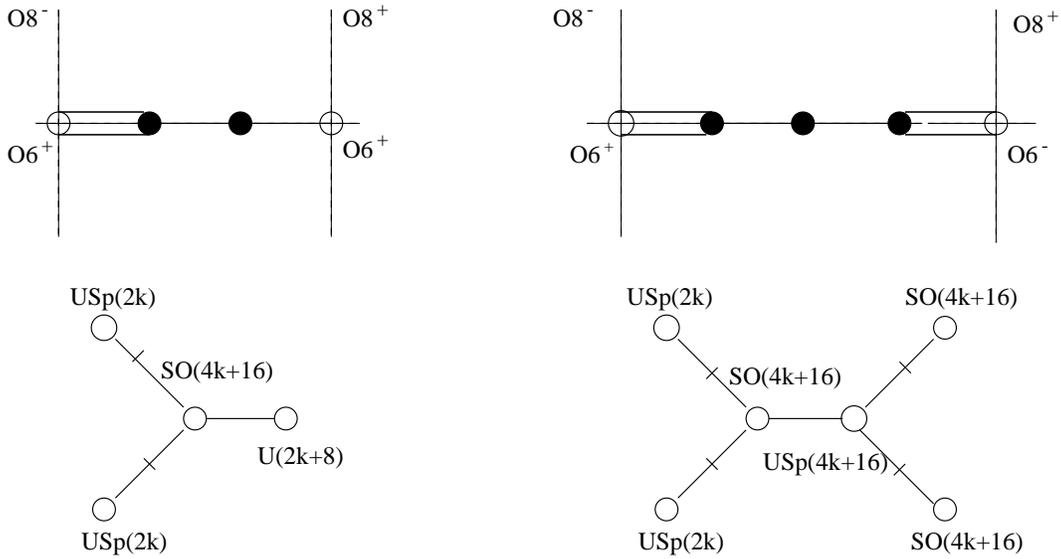}} \par}

\caption{\protect\( D_{4}\protect \) and \protect\( D_{5}\protect \) theories constructed
with the use of \protect\( O8^{+}\protect \).\label{oeightplus}}
\end{figure}

A second obvious generalization involves the introduction of an \( O8^{+} \)
plane. It has charge \( +8 \), therefore if we have both \( O8^{-} \) and
\( O8^{+} \) in a compact model there is no need for D8-branes. The discussion
about the properties of \( O8-ON^{0} \) planes is invariant under the simultaneous
change of sign of the charge of the \( O8 \) and \( O6 \) planes. This means
that there are two consistent configurations of D-branes living at the intersection
of an \( O8^{+} \) and an \( ON^{0} \) plane. The first one has an \( O6^{+} \)
plane, no splitting of branes, \( U \)-type gauge groups and a hypermultiplet
from the twisted states. The second one has an \( O6^{-} \) plane, a splitting
with \( SO \)-type gauge groups and a tensor multiplet from the twisted states.
In the case of a \( D_{n} \) singularity, we obtain standard D-type Dynkin
diagrams for \( n \) odd, and extended (affine) D-type Dynkin diagrams for
\( n \) even. This is the opposite of what we found when the two \( O8 \)
planes had both the same negative charge. The \( D_{4} \) and \( D_{5} \)
cases are depicted in figure \ref{oeightplus}.

We discussed several examples. They are complete in the sense that they cover
all the possible configurations which contain \( ON \), \( O8 \) and \( O6 \)
planes. By combining in a different way the various players of this game, we
can construct many other compact and non-compact models. The rules we gave in
the previous section for determining the world-volume theory will always produce
a consistent anomaly-free six-dimensional theory with a non-trivial fixed point.
We will not discuss the applications of what we said to all the rich aspects
of the six-dimensional physics, since the general philosophy of the brane construction
for six dimensional theories was already presented in \cite{hzsix}.

\section*{Acknowledgements}

A. H. would like to thank discussions with Jacques Distler and Bo Feng. A. H.
Would also like to thank the Theoretical Physics Division in CERN and the Institute
for Advanced Study in Jerusalem for their kind hospitality while various stages
of this work were done. A. Z. would like to thank the Center for Theoretical
Physics at MIT for kind hospitality during the early stages of this work. The
work of A. H. was supported in part by the U.S. Department of Energy under contract
\#DE-FC02-94ER40818. 

{\small \bibliographystyle{utphys}
\bibliography{orientifolds}

\providecommand{\href}[2]{#2}\begingroup\raggedright\begin{thebibliography}{10}

\bibitem{asadjosh}
J.~Erlich, A.~Hanany, and A.~Naqvi, ``Marginal deformations from branes,''
  \href{http://xxx.lanl.gov/abs/hep-th/9902118}{{\tt hep-th/9902118}}.

\bibitem{witten}
E.~Witten, ``Baryons and branes in anti-de {Sitter} space,'' {\em JHEP} {\bf
  07} (1998) 006, \href{http://xxx.lanl.gov/abs/hep-th/9805112}{{\tt
  hep-th/9805112}}.

\bibitem{uranga}
A.~M. Uranga, ``Towards mass deformed {N=4 SO(n) and Sp(k)} gauge theories from
  brane configurations,'' {\em Nucl. Phys.} {\bf B526} (1998) 241,
  \href{http://xxx.lanl.gov/abs/hep-th/9803054}{{\tt hep-th/9803054}}.

\bibitem{hori}
K.~Hori, ``Consistency condition for five-brane in {M} theory on {R$^5$ /
  Z$_2$} orbifold,'' {\em Nucl. Phys.} {\bf B539} (1999) 35,
  \href{http://xxx.lanl.gov/abs/hep-th/9805141}{{\tt hep-th/9805141}}.

\bibitem{Sen}
A.~Sen, ``Stable {nonBPS} states in string theory,'' {\em JHEP} {\bf 06} (1998)
  007, \href{http://xxx.lanl.gov/abs/hep-th/9803194}{{\tt hep-th/9803194}}.

\bibitem{senS}
A.~Sen, ``Stable {nonBPS bound states of BPS D-branes},'' {\em JHEP} {\bf 08}
  (1998) 010, \href{http://xxx.lanl.gov/abs/hep-th/9805019}{{\tt
  hep-th/9805019}}.

\bibitem{kapu}
A.~Kapustin, ``{D}$_n$ quivers from branes,''
  \href{http://xxx.lanl.gov/abs/hep-th/9806238}{{\tt hep-th/9806238}}.

\bibitem{hw}
A.~Hanany and E.~Witten, ``{Type IIB} superstrings, {BPS} monopoles, and
  three-dimensional gauge dynamics,'' {\em Nucl. Phys.} {\bf B492} (1997)
  152--190, \href{http://xxx.lanl.gov/abs/hep-th/9611230}{{\tt
  hep-th/9611230}}.

\bibitem{pz}
M.~Porrati and A.~Zaffaroni, ``M theory origin of mirror symmetry in
  three-dimensional gauge theories,'' {\em Nucl. Phys.} {\bf B490} (1997)
  107--120, \href{http://xxx.lanl.gov/abs/hep-th/9611201}{{\tt
  hep-th/9611201}}.

\bibitem{hzfour}
A.~Hanany and A.~Zaffaroni, ``On the realization of chiral four-dimensional
  gauge theories using branes,'' {\em JHEP} {\bf 05} (1998) 001,
  \href{http://xxx.lanl.gov/abs/hep-th/9801134}{{\tt hep-th/9801134}}.

\bibitem{hstrassu}
A.~Hanany, M.~J. Strassler, and A.~M. Uranga, ``Finite theories and marginal
  operators on the brane,'' {\em JHEP} {\bf 06} (1998) 011,
  \href{http://xxx.lanl.gov/abs/hep-th/9803086}{{\tt hep-th/9803086}}.

\bibitem{hu}
A.~Hanany and A.~M. Uranga, ``Brane boxes and branes on singularities,'' {\em
  JHEP} {\bf 05} (1998) 013, \href{http://xxx.lanl.gov/abs/hep-th/9805139}{{\tt
  hep-th/9805139}}.

\bibitem{karch}
I.~Brunner and A.~Karch, ``Branes and six-dimensional fixed points,'' {\em
  Phys. Lett.} {\bf B409} (1997) 109--116,
  \href{http://xxx.lanl.gov/abs/hep-th/9705022}{{\tt hep-th/9705022}}.

\bibitem{hzsix}
A.~Hanany and A.~Zaffaroni, ``Branes and six-dimensional supersymmetric
  theories,'' {\em Nucl. Phys.} {\bf B529} (1998) 180,
  \href{http://xxx.lanl.gov/abs/hep-th/9712145}{{\tt hep-th/9712145}}.

\bibitem{karchbr}
I.~Brunner and A.~Karch, ``Branes at orbifolds versus {Hanany Witten} in six-
  dimensions,'' {\em JHEP} {\bf 03} (1998) 003,
  \href{http://xxx.lanl.gov/abs/hep-th/9712143}{{\tt hep-th/9712143}}.

\bibitem{hm}
M.~R. Douglas and G.~Moore, ``{D}-branes, quivers, and {ALE} instantons,''
  \href{http://xxx.lanl.gov/abs/hep-th/9603167}{{\tt hep-th/9603167}}.

\bibitem{gp}
E.~G. Gimon and J.~Polchinski, ``Consistency conditions for orientifolds and d
  manifolds,'' {\em Phys. Rev.} {\bf D54} (1996) 1667--1676,
  \href{http://xxx.lanl.gov/abs/hep-th/9601038}{{\tt hep-th/9601038}}.

\bibitem{senT}
A.~Sen, ``Duality and orbifolds,'' {\em Nucl. Phys.} {\bf B474} (1996)
  361--378, \href{http://xxx.lanl.gov/abs/hep-th/9604070}{{\tt
  hep-th/9604070}}.

\bibitem{leigh}
M.~Berkooz {\em et.~al.}, ``Anomalies, dualities, and topology of d = 6 {N=1}
  superstring vacua,'' {\em Nucl. Phys.} {\bf B475} (1996) 115--148,
  \href{http://xxx.lanl.gov/abs/hep-th/9605184}{{\tt hep-th/9605184}}.

\bibitem{p}
J.~Polchinski, ``Tensors from {K3} orientifolds,'' {\em Phys. Rev.} {\bf D55}
  (1997) 6423--6428, \href{http://xxx.lanl.gov/abs/hep-th/9606165}{{\tt
  hep-th/9606165}}.

\bibitem{intblum1}
J.~D. Blum and K.~Intriligator, ``Consistency conditions for branes at orbifold
  singularities,'' {\em Nucl. Phys.} {\bf B506} (1997) 223,
  \href{http://xxx.lanl.gov/abs/hep-th/9705030}{{\tt hep-th/9705030}}.

\bibitem{intblum2}
J.~D. Blum and K.~Intriligator, ``New phases of string theory and 6-d {RG}
  fixed points via branes at orbifold singularities,'' {\em Nucl. Phys.} {\bf
  B506} (1997) 199, \href{http://xxx.lanl.gov/abs/hep-th/9705044}{{\tt
  hep-th/9705044}}.

\bibitem{gj}
E.~G. Gimon and C.~V. Johnson, ``K3 orientifolds,'' {\em Nucl. Phys.} {\bf
  B477} (1996) 715--745, \href{http://xxx.lanl.gov/abs/hep-th/9604129}{{\tt
  hep-th/9604129}}.

\bibitem{charge}
N.~Evans, C.~V. Johnson, and A.~D. Shapere, ``Orientifolds, branes, and duality
  of 4-d gauge theories,'' {\em Nucl. Phys.} {\bf B505} (1997) 251,
  \href{http://xxx.lanl.gov/abs/hep-th/9703210}{{\tt hep-th/9703210}}.

\bibitem{vaoo}
H.~Ooguri and C.~Vafa, ``Two-dimensional black hole and singularities of {CY}
  manifolds,'' {\em Nucl. Phys.} {\bf B463} (1996) 55--72,
  \href{http://xxx.lanl.gov/abs/hep-th/9511164}{{\tt hep-th/9511164}}.

\bibitem{kut}
D.~Kutasov, ``Orbifolds and solitons,'' {\em Phys. Lett.} {\bf B383} (1996)
  48--53, \href{http://xxx.lanl.gov/abs/hep-th/9512145}{{\tt hep-th/9512145}}.

\bibitem{deboer}
J.~de~Boer, K.~Hori, H.~Ooguri, and Y.~Oz, ``Mirror symmetry in
  three-dimensional gauge theories, quivers and {D}-branes,'' {\em Nucl. Phys.}
  {\bf B493} (1997) 101--147,
  \href{http://xxx.lanl.gov/abs/hep-th/9611063}{{\tt hep-th/9611063}}.

\bibitem{horivafa}
K.~Hori, H.~Ooguri, and C.~Vafa, ``{NonAbelian conifold transitions and N=4}
  dualities in three- dimensions,'' {\em Nucl. Phys.} {\bf B504} (1997) 147,
  \href{http://xxx.lanl.gov/abs/hep-th/9705220}{{\tt hep-th/9705220}}.

\bibitem{intseib}
K.~Intriligator and N.~Seiberg, ``Mirror symmetry in three-dimensional gauge
  theories,'' {\em Phys. Lett.} {\bf B387} (1996) 513--519,
  \href{http://xxx.lanl.gov/abs/hep-th/9607207}{{\tt hep-th/9607207}}.

\bibitem{fkpz}
S.~Ferrara, A.~Kehagias, H.~Partouche, and A.~Zaffaroni, ``Membranes and
  five-branes with lower supersymmetry and their {AdS} supergravity duals,''
  {\em Phys. Lett.} {\bf B431} (1998) 42--48,
  \href{http://xxx.lanl.gov/abs/hep-th/9803109}{{\tt hep-th/9803109}}.

\bibitem{gomis}
J.~Gomis, ``Anti-de {Sitter} geometry and strongly coupled gauge theories,''
  {\em Phys. Lett.} {\bf B435} (1998) 299,
  \href{http://xxx.lanl.gov/abs/hep-th/9803119}{{\tt hep-th/9803119}}.

\bibitem{sethi}
S.~Sethi, ``A relation between {N=8} gauge theories in three-dimensions,'' {\em
  JHEP} {\bf 11} (1998) 003, \href{http://xxx.lanl.gov/abs/hep-th/9809162}{{\tt
  hep-th/9809162}}.

\bibitem{kapuber}
M.~Berkooz and A.~Kapustin, ``New {IR} dualities in supersymmetric gauge theory
  in three- dimensions,'' \href{http://xxx.lanl.gov/abs/hep-th/9810257}{{\tt
  hep-th/9810257}}.

\bibitem{list1}
E.~G. Gimon and M.~Gremm, ``A note on brane boxes at finite string coupling,''
  {\em Phys. Lett.} {\bf B433} (1998) 318,
  \href{http://xxx.lanl.gov/abs/hep-th/9803033}{{\tt hep-th/9803033}}.

\bibitem{list2}
L.~Randall, Y.~Shirman, and R.~von Unge, ``Brane boxes: Bending and beta
  functions,'' {\em Phys. Rev.} {\bf D58} (1998) 105005,
  \href{http://xxx.lanl.gov/abs/hep-th/9806092}{{\tt hep-th/9806092}}.

\bibitem{list3}
R.~G. Leigh and M.~Rozali, ``Brane boxes, anomalies, bending and tadpoles,''
  {\em Phys. Rev.} {\bf D59} (1999) 026004,
  \href{http://xxx.lanl.gov/abs/hep-th/9807082}{{\tt hep-th/9807082}}.

\bibitem{list4}
A.~Karch, D.~Lust, and A.~Miemiec, ``N=1 supersymmetric gauge theories and
  supersymmetric three cycles,''
  \href{http://xxx.lanl.gov/abs/hep-th/9810254}{{\tt hep-th/9810254}}.

\bibitem{witten2}
E.~Witten, ``Solutions of four-dimensional field theories via {M} theory,''
  {\em Nucl. Phys.} {\bf B500} (1997) 3,
  \href{http://xxx.lanl.gov/abs/hep-th/9703166}{{\tt hep-th/9703166}}.

\bibitem{poppitz}
J.~Lykken, E.~Poppitz, and S.~P. Trivedi, ``Chiral gauge theories from
  {D}-branes,'' {\em Phys. Lett.} {\bf B416} (1998) 286,
  \href{http://xxx.lanl.gov/abs/hep-th/9708134}{{\tt hep-th/9708134}}.

\bibitem{hevariste}
A.~Hanany and Y.-H. He, ``{NonAbelian} finite gauge theories,''
  \href{http://xxx.lanl.gov/abs/hep-th/9811183}{{\tt hep-th/9811183}}.

\bibitem{eva}
S.~Kachru and E.~Silverstein, ``4-d conformal theories and strings on
  orbifolds,'' {\em Phys. Rev. Lett.} {\bf 80} (1998) 4855,
  \href{http://xxx.lanl.gov/abs/hep-th/9802183}{{\tt hep-th/9802183}}.

\bibitem{augusto1}
A.~Sagnotti, ``Open strings and their symmetry groups,''. Talk presented at the
  Cargese Summer Institute on Non- Perturbative Methods in Field Theory,
  Cargese, Italy, Jul 16-30, 1987.

\bibitem{augusto2}
M.~Bianchi and A.~Sagnotti, ``On the systematics of open string theories,''
  {\em Phys. Lett.} {\bf B247} (1990) 517--524.

\bibitem{augusto3}
M.~Bianchi and A.~Sagnotti, ``Twist symmetry and open string {Wilson} lines,''
  {\em Nucl. Phys.} {\bf B361} (1991) 519--538.

\bibitem{hzfour2}
A.~Hanany and A.~Zaffaroni, ``Chiral symmetry from type {IIA} branes,'' {\em
  Nucl. Phys.} {\bf B509} (1998) 145,
  \href{http://xxx.lanl.gov/abs/hep-th/9706047}{{\tt hep-th/9706047}}.

\bibitem{sagnotti}
A.~Sagnotti, ``A note on the green-schwarz mechanism in open string theories,''
  {\em Phys. Lett.} {\bf B294} (1992) 196--203,
  \href{http://xxx.lanl.gov/abs/hep-th/9210127}{{\tt hep-th/9210127}}.

\bibitem{wpol}
J.~Polchinski and E.~Witten, ``Evidence for heterotic - type {I} string
  duality,'' {\em Nucl. Phys.} {\bf B460} (1996) 525--540,
  \href{http://xxx.lanl.gov/abs/hep-th/9510169}{{\tt hep-th/9510169}}.

\bibitem{am}
P.~S. Aspinwall and D.~R. Morrison, ``Point - like instantons on {K3}
  orbifolds,'' {\em Nucl. Phys.} {\bf B503} (1997) 533,
  \href{http://xxx.lanl.gov/abs/hep-th/9705104}{{\tt hep-th/9705104}}.

\end{thebibliography}\endgroup
}{\small \par}

\end{document}